\documentclass[%
 reprint,
superscriptaddress,
 amsmath,amssymb,
 aps,
pra,
]{revtex4-1}

\usepackage{amsmath}
\usepackage{amssymb}
\usepackage{url}
\usepackage{placeins}
\usepackage{textcomp}
\usepackage{graphicx}
\usepackage[usenames,dvipsnames]{color}
\usepackage{soul}

\usepackage{soul}   
\setstcolor{red} 
\setul{}{1.5pt}  

\newcommand{\mb}[1]{\mathbf{#1}} 

\begin{document}

\title{Optimization of Anisotropic Photonic Density of States for Raman Cooling of Solids}

\author{Yin-Chung Chen}

\affiliation{Department of Mechanical Science and Engineering, University of Illinois at Urbana-Champaign, Urbana, IL 61801, USA}

\author{Indronil Ghosh}

\affiliation{Department of Mechanical Science and Engineering, University of Illinois at Urbana-Champaign, Urbana, IL 61801, USA}

\author{Andr\'{e} Schleife}

\affiliation{Department of Materials Science and Engineering, University of Illinois at Urbana-Champaign, Urbana, IL 61801, USA}

\author{P. Scott Carney}
\affiliation{Institute of Optics, University of Rochester, Rochester, NY 14627, USA}

\author{Gaurav Bahl}
\email[]{bahl@illinois.edu}
\affiliation{Department of Mechanical Science and Engineering, University of Illinois at Urbana-Champaign, Urbana, IL 61801, USA}

\begin{abstract}
Optical refrigeration of solids holds tremendous promise for applications in thermal management. It can be achieved through multiple mechanisms including inelastic anti-Stokes Brillouin and Raman scattering. However, engineering of these mechanisms remains relatively unexplored. The major challenge lies in the natural unfavorable imbalance in transition rates for Stokes and anti-Stokes scattering.  We consider the influence of anisotropic photonic density of states on Raman scattering and derive expressions for cooling in such systems.  We demonstrate optimization of the Raman cooling figure of merit considering all possible orientations for the material crystal and two example photonic crystals. We find that the anisotropic description of the photonic DoS and the optimization process is necessary to obtain the best Raman cooling efficiency for systems having lower symmetry. This general result applies to a wide array of other laser cooling methods in the presence of anisotropy.
\end{abstract}

\maketitle

\section{Introduction}

\begin{figure*}[!ht]
\begin{center}
\includegraphics[trim=0cm 0cm 0cm 0cm, clip, width=0.8\textwidth]{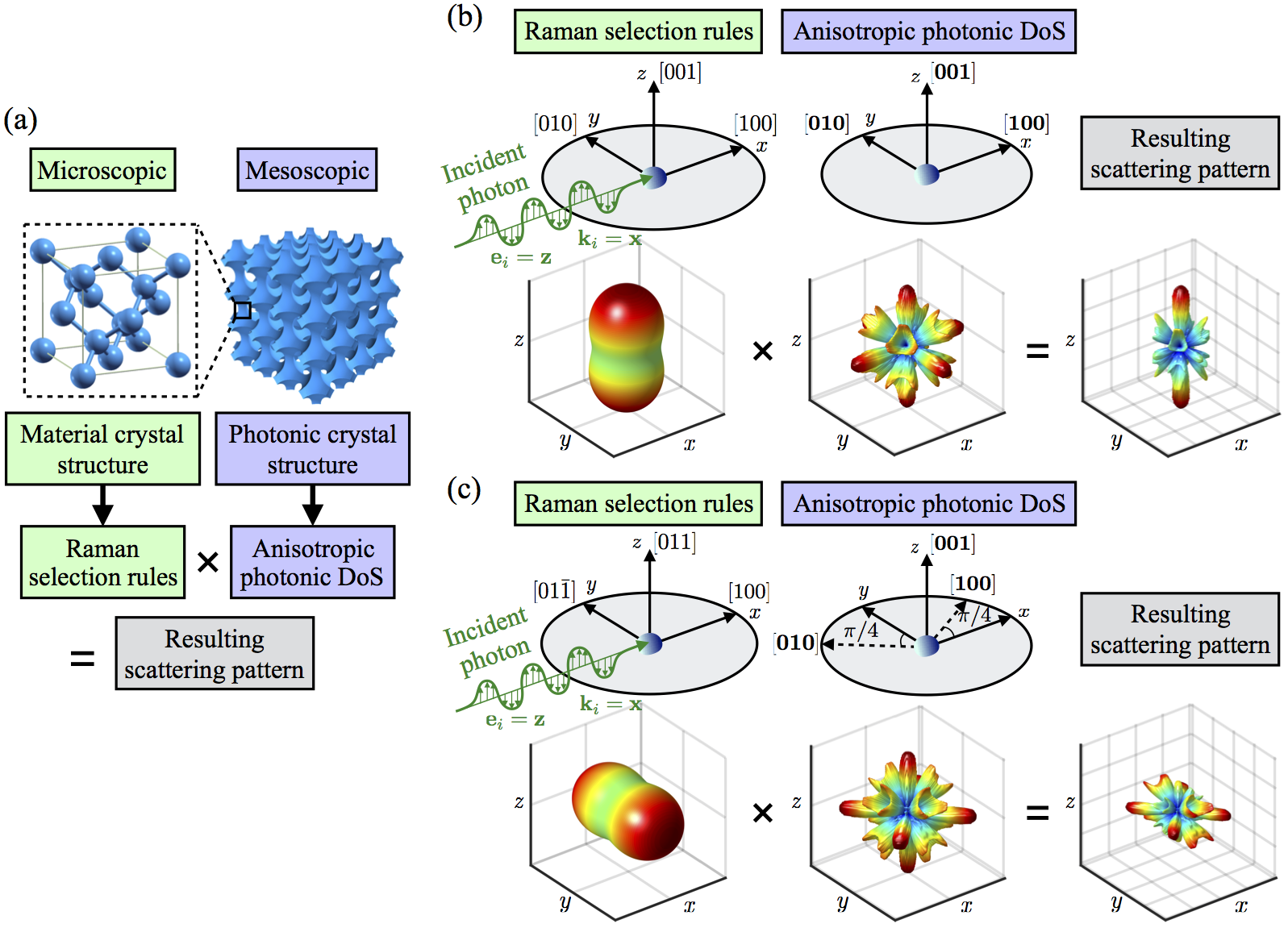}
\caption{Illustration of how Raman selection rules (Eq. (\ref{eq:RI})) and anisotropic photonic DoS (Eq. (\ref{eq:AIDoSC})) affect the total Raman scattering pattern. \textbf{(a)} The material crystal structure (example of silicon, diamond crystal structure) and photonic crystal structure (example of simple cubic with air spheres) used in this paper. \textbf{(b)} An example of a resulting Raman scattering pattern for a chosen set of material crystal and photonic crystal orientation with respect to an incident photon with a fixed propagation direction ($\mb{k}_i$) and a fixed polarization ($\mb{e}_i$). \textbf{(c)} Changes in the Raman scattering pattern when both the material and the photonic crystal structures in \textbf{(b)} are rotated while keeping the incident polarization fixed ($\mb{e}_i$).}
\label{fig:RTDoS}
\end{center}
\end{figure*}

Since the first proposal of the concept of optical cooling by Peter Pringsheim in 1929 \cite{pringsheim1929zwei}, various experimental demonstrations have been produced in atomic gases \cite{hansch1975cooling,dalibard1985dressed,aspect1988laser} and in solids \cite{epstein1995observation,seletskiy2010laser,seletskiy2012cryogenic,melgaard2013optical,melgaard2014identification,zhang2013laser}. Laser cooling of \textit{solids} in particular, can be achieved by anti-Stokes photon up-conversion processes in which thermal phonons are annihilated \cite{PhysRevLett.92.247403,sheik2009laser,ding2012anti}. The most common approach employs anti-Stokes fluorescence produced by specific electronic transitions in certain rare-earth elements (e.g. Yb$^{3+}$, Er$^{3+}$, Tm$^{3+}$)\cite{epstein1995observation,seletskiy2010laser,seletskiy2012cryogenic,melgaard2013optical,melgaard2014identification}. 
An alternative approach uses strong exciton-phonon interaction in a polar material to achieve cooling via photoluminescence \cite{zhang2013laser}. Furthermore, in order to minimize non-radiative recombination processes, only direct band gap materials are suitable. Even though these techniques for cooling solids have been remarkably successful \cite{melgaard2013optical,melgaard2016solid}, the reliance on specific materials restricts their broad applicability.

Inelastic light scattering, which also provides anti-Stokes photon up-conversion and has fewer restrictions on material properties, has separately emerged as an alternative method for laser cooling of solids. The fundamental challenge for cooling using inelastic light scattering (e.g. Brillouin or Raman scattering) lies in the natural imbalance between the Stokes and the anti-Stokes scattering rates. In a homogeneous material the spontaneous scattering rate for the Stokes process is proportional to $n_0+1$ while the scattering rate for the anti-Stokes process is proportional to $n_0$, where $n_0=[\exp(\hbar \omega_0/k_BT)-1]^{-1}$ is the Bose-Einstein distribution and $\omega_0$ is the phonon frequency. For optical phonons in silicon, for example, this ratio of Stokes transition rate to anti-Stokes transition rate is about 10 to 1 at room temperature.
To overcome this challenge researchers have attempted to engineer either the electronic properties \cite{khurgin2006band,ding2012anti,zhang2013laser,zhang2016resolved} or the optical properties \cite{bahl2012observation,chen2015raman,chen2016brillouin, kim2017role} of the system. For instance, successful engineering of electronic properties via excitonic resonance has led to the demonstration of Raman cooling of a longitudinal optical (LO) phonon mode in ZnTe nanobelts \cite{zhang2016resolved}, in which only the anti-Stokes scattering is in excitonic resonance.

The engineering of optical properties, on the other hand, focuses on enhancing the available photonic states \cite{purcell1946spontaneous, gaponenko2002effects} for only anti-Stokes scattered photons.
Using this concept with optical resonators resulted in the first demonstration of Brillouin cooling of an acoustic phonon mode \cite{bahl2012observation}, and the understanding that the cooling efficiency is shaped by the photonic density of states (DoS) of the optical modes \cite{kim2017role}.

In addition to dealing with only a single resonant mode of the system, we further expand this method to incorporate the engineered DoS of all the photonic modes of the system in the context of Raman scattering \cite{chen2015raman}. By combining the effect of Raman selection rules and the photonic DoS, we showed that net Raman cooling can be potentially achieved in a system with low absorption.
However, the photonic DoS in the calculation was assumed to be isotropic, which as we will see in the present study is insufficient in dealing with certain photonic systems. Furthermore, the fundamental problem of analyzing Raman scattering in a system considering the full effect of anisotropic photonic DoS has not previously been studied. In this study we produce for the first time an explicit expression for calculating anisotropic DoS in a photonic engineered system and analyze the effect of anisotropic DoS on Raman scattering. We apply this new knowledge to the specific problem of Raman laser cooling in solids and perform a full optimization of the cooling process.

The basic concept of using anisotropic photonic DoS to engineer Raman scattering is illustrated in Fig. \ref{fig:RTDoS}. The scattering efficiency is strongly influenced by structure at two scales. First, the crystal structure at the microscopic, molecular scale determines available vibrational states and the resulting Raman transitions. For a selected incident light polarization, the response at this microscopic scale will determine the Raman scattering intensity in all directions (for example: left panel of Fig. \ref{fig:RTDoS}b). If we change the material crystal orientation with respect to a fixed incident polarization, the resulting scattering pattern will also change accordingly (left panel of Fig. \ref{fig:RTDoS}c).

Second, at the mesoscopic scale, the photonic structure (e.g. photonic crystal) determines the available states for light in the material or the anisotropic photonic DoS (for example: middle panel of Fig. \ref{fig:RTDoS}b) of the system. If the principal axes of the photonic crystal are rotated, i.e. the patterning of air-spheres is changed, the photonic DoS pattern will also be rotated (middle panel of Fig. \ref{fig:RTDoS}c).

These two effects at the microscale and mesoscale together determine the Raman efficiency and the resulting scattering pattern is given by the product of the anisotropic DoS and the Raman scattering intensity (right panel of Fig. \ref{fig:RTDoS}b and \ref{fig:RTDoS}c). By arranging the relative orientation of the photonic crystal and the material crystal we can thus obtain the best resulting Raman scattering efficiency for various Raman applications. In this paper, we aim to exploit this combined effect at two scales and propose how to optimize the relative orientation for maximizing Raman cooling efficiency.

To demonstrate our approach, we use crystalline silicon as our material crystal since its highly symmetric structure and well-known properties simplify the calculation. In addition, we choose two different photonic crystals with different symmetries to emphasis the effect of anisotropy on Raman cooling efficiency.

To standardize coordinate systems and crystal axes, we use $x$, $y$, and $z$ to denote a fixed coordinate system, which is also used to define the polarization direction of the incident light. Miller indices with bold numbers (e.g. $[\textbf{100}]$, $[\textbf{010}]$) are used to denote directions with respect to the photonic crystal, and the usual Miller indices (e.g. $[100]$, $[11\bar{2}]$) denote directions with respect to the material crystal.

The paper is organized as follows. In section \ref{sec:ADoS} we review the theory of Raman scattering and produce an expression of anisotropic photonic DoS. In section \ref{sec:RSR} we discuss the effect of Raman selection rules on the Raman scattering intensity. Finally, in section \ref{sec:FoM} we discuss the application to Raman laser cooling.

\section{Raman scattering and anisotropic DoS}\label{sec:ADoS}

In this section, we will provide an explicit expression for calculating anisotropic DoS and discuss how it affects the Raman transition rate. Later, we will directly employ the anisotropic DoS in the calculation of the Raman scattering efficiency, which is directly proportional to the Raman transition rate.

In a \textit{bulk homogeneous} medium, the total transition rate for Raman scattering with pump frequency $\omega_i$ and phonon frequency $\omega_0$ is given by Fermi's golden rule \cite{loudon1963theory,loudon1964raman}:
\begin{align}\label{eq:TRSI}
W_{S}= \frac{1}{V}\sum_{\mb{k}} n_i (n_0+1)|\mathcal{M}_{S}(\mb{k})|^2 \delta(\omega_i-\omega_0-\omega(\mb{k}))
\end{align}
for the Stokes process and
\begin{align}\label{eq:TRASI}
W_{AS}= \frac{1}{V}\sum_{\mb{k}} n_i n_0 |\mathcal{M}_{AS}(\mb{k})|^2 \delta(\omega_i+\omega_0-\omega(\mb{k}))
\end{align}
for the anti-Stokes process, where $V$ is the volume of the system, $n_i$ is the number of pump photons, and $n_0=[\exp(\hbar\omega_0/k_BT)-1]^{-1}$ is the equilibrium phonon occupation number at temperature $T$. The delta functions in the above equations enforce energy conservation, where $\omega_i-\omega_0=\omega_{S}$ is the Stokes frequency, $\omega_i+\omega_0=\omega_{AS}$ is the anti-Stokes frequency, and $\omega(\mb{k})$ is the photon-dispersion relation of the system. $\mathcal{M}_{S}$ and $\mathcal{M}_{AS}$ are summations over all possible intermediate transition matrix elements for the Stokes and anti-Stokes process, which include both electron-photon interaction and electron-phonon interaction \cite{loudon1963theory}. They are in general functions of both frequencies and wave vectors of the photons and phonons participating in scattering process as well as the electron state energies for the virtual transitions. The summations in Eqs. (\ref{eq:TRSI}) and (\ref{eq:TRASI}) should also be over all possible final states, which would include both final photon and final phonon states labeled by their respective wave vectors $\mb{k}$ and $\mb{q}$. 
However, since we consider only first order non-resonant Raman scattering in this paper, the only phonon modes that participate in the scattering process are the zone center Raman active phonon modes \cite{yu1996fundamentals}. We can therefore neglect the dependence of $\mathcal{M}_{S}$ and $\mathcal{M}_{AS}$ on frequency and phonon momentum $\mb{q}$, and sum over only all final photon states labeled by $\mb{k}$. In most experiments, the pump frequency is much greater than the phonon frequency $\omega_i\gg\omega_0$, which allows us to approximate $|\mathcal{M}_{S}(\mb{k})|\approx |\mathcal{M}_{AS}(\mb{k})|$ \cite{yu1996fundamentals}.
As we discuss later in section \ref{sec:RSR}, the dependence of $\mathcal{M}_{S}$ and $\mathcal{M}_{AS}$ on $\mb{k}$ comes only from the direction of the final photon state $\hat{\Omega}(\mb{k})$. We can thus rewrite them as $\mathcal{M}_{S}(\hat{\Omega}(\mb{k}))$ and $\mathcal{M}_{AS}(\hat{\Omega}(\mb{k}))$.

In a bulk homogeneous medium, the dependence of the transition rate on the direction of scattered light only enters via $\mathcal{M}_{S}$ and $\mathcal{M}_{AS}$ since the photon-dispersion relation is isotropic and given by $\omega(\mb{k})=c|\mb{k}|$. If we convert the sum into integral in Eq. (\ref{eq:TRSI}), we have
\begin{align}\label{eq:TRSIInt}
W_{S}= \iint \frac{k^2dk \  d\Omega}{(2\pi)^3} n_i (n_0+1)|\mathcal{M}_{S}(\hat{\Omega})|^2 \delta(\omega_S-ck),
\end{align}
where a spherical coordinate system is used, and $\mb{k}=(k,\hat{\Omega})$. Here $\mathcal{M}_{S}$ is only a function of direction $\hat{\Omega}$ and is independent of the magnitude $k$. Thus, we can write down the transition rate observed over a unit solid angle as
\begin{align}
\frac{dW_{S}}{d\Omega}=n_i (n_0+1)|\mathcal{M}_{S}(\hat{\Omega})|^2 D_0(\omega_S),
\end{align}
where
\begin{align}\label{eq:IsoDoS}
D_0(\omega)=\int \frac{k^2dk}{(2\pi)^3}  \delta(\omega-ck)=\frac{\omega^2}{(2\pi c)^3}
\end{align}
is the isotropic photonic DoS per unit solid angle and is independent of $\hat{\Omega}$ \cite{gaponenko2002effects}. In this case, the photonic density of states factor can be taken out of the integration over solid angle in the calculation of total scattering rate, as demonstrated in \cite{chen2015raman}. 

\begin{figure*}[!ht]
\begin{center}
\includegraphics[trim=0cm 0cm 0cm 0cm, clip, width=0.8\textwidth]{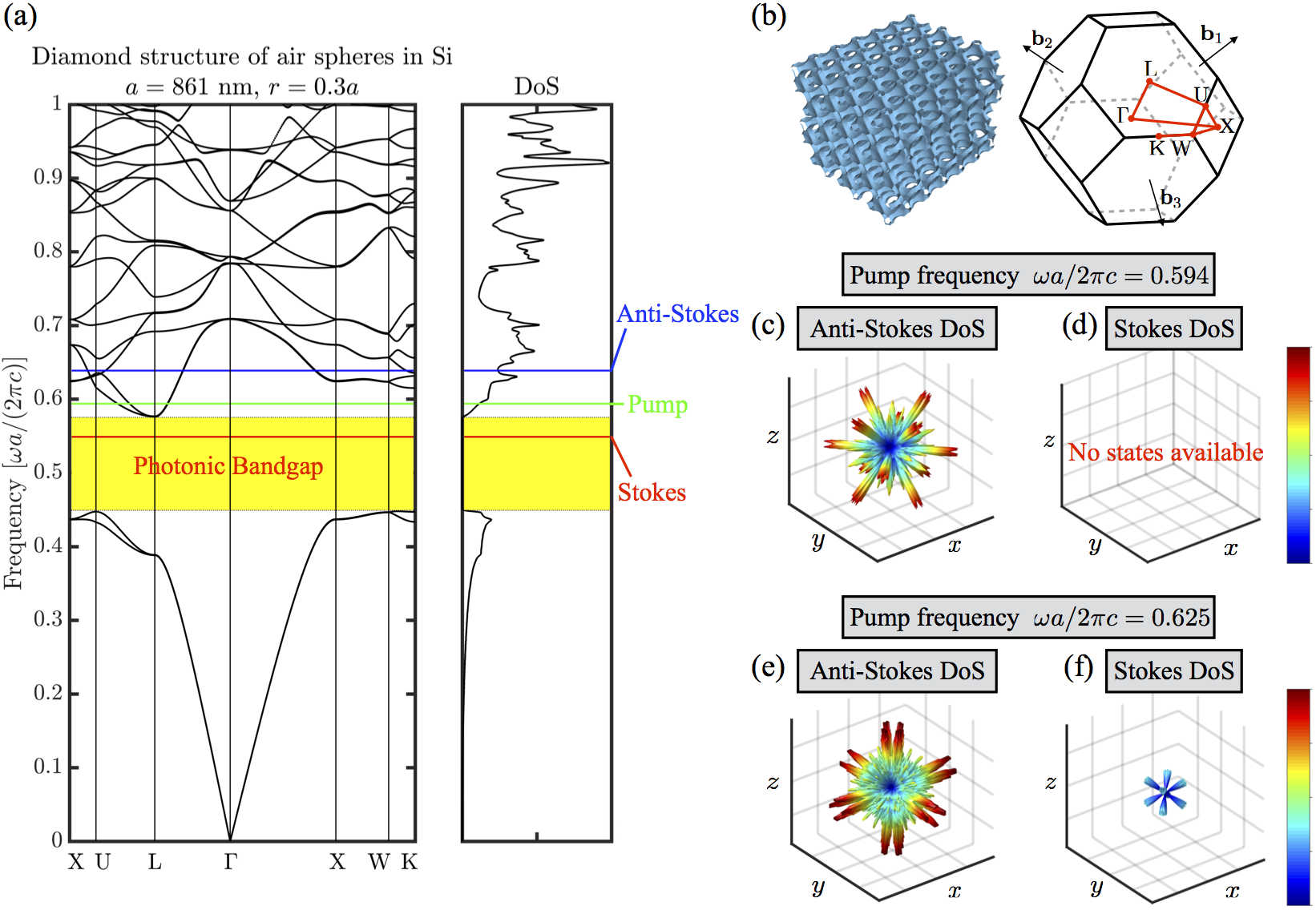}
\caption{Examples of anisotropic photonic DoS for a diamond photonic crystal of air spheres of radius $r=0.3a$ patterned in silicon, where $a$ is the lattice constant. \textbf{(a)} Band structure and the isotropic DoS of the system. \textbf{(b)} Structure of the photonic crystal and the $k$-point path in the Brillouin zone of the photonic crystal. We set the pump at $\omega a/2\pi c=0.594$, and use the zone-center phonon energy for silicon (519 cm$^{-1}$), corresponding to a non-dimensional frequency shift of 0.045 here ($a=861$ nm). This leads to an anti-Stokes frequency at $\omega a/2\pi c=0.639$, and a Stokes frequency at $\omega a/2\pi c=0.549$. If the pump is shifted to $\omega a/2\pi c=0.625$, the anti-Stokes frequency is shifted to $\omega a/2\pi c=0.670$, and the Stokes frequency to $\omega a/2\pi c=0.580$.  Computed anisotropic DoS \textbf{(c)} at the anti-Stokes frequency and \textbf{(d)} at the Stokes frequency when pumped at $\omega a/2\pi c=0.594$, and \textbf{(e)} at the anti-Stokes frequency and \textbf{(f)} at the Stokes frequency when pumped at $\omega a/2\pi c=0.625$. All the values of DoS are normalized by the maximum value among the four DoS plots. For anisotropic DoS at multiple frequencies along the band structure, see the supplementary video.}
\label{fig:BandDia}
\end{center}
\end{figure*}

\vspace{10pt}

In a \textit{photonic crystal} system with infinite periodicity, the system is amenable to the standard Bloch treatment and only the modes in the first Brillouin zone are significant. Eq. (\ref{eq:TRSI}) (similarly for Eq. (\ref{eq:TRASI})) then becomes
\begin{align}\label{eq:TRSPC}
W_{S}= \sum_\alpha &\int_{\text{FBZ}} \frac{d\mb{k}}{(2\pi)^3} n_i (n_0+1)|\mathcal{M}_{S}(\hat{\Omega}_\alpha(\mb{k}))|^2 \notag \\
	&\quad\times\delta(\omega_S-\omega_\alpha(\mb{k})),
\end{align}
where we have converted the sum over $\mb{k}$ in Eq. (\ref{eq:TRSI}) into an integral over all possible wave vectors in the first Brillouin zone. Here we also sum over all bands labeled by $\alpha$. Each wave vector $\mb{k}$ is now associated with more than one eigenfrequency. It is therefore important to also label the direction of the wave vector corresponding to each band $\hat{\Omega}_\alpha(\mb{k})$. In this case both the matrix elements $\mathcal{M}_{S}$ and $\mathcal{M}_{AS}$ and the delta-functions depend on the direction of the scattered light. Due to the anisotropic dispersion relation, we cannot separate the integral over $\mb{k}$ into integrals over $dk$ and $d\Omega$ as we did in the isotropic case. However, if we focus our attention on the transition rate for light scattered into a differential solid angle $d\Omega$ around the direction of interest $\hat{\Omega}$, we can write
\begin{align}\label{eq:TRSAI}
\frac{dW_{S}}{d\Omega}&=\sum_\alpha  \int_{\text{FBZ}} \frac{d\mb{k}}{(2\pi)^3} n_i (n_0+1)|\mathcal{M}_{S}(\hat{\Omega})|^2  \notag \\
&\qquad\qquad \times\delta(\omega_S-\omega_\alpha(\mb{k}))\delta(\hat{\Omega}-\hat{\Omega}_\alpha(\mb{k}))\notag \\
&=n_i (n_0+1)|\mathcal{M}_{S}(\hat{\Omega})|^2 D(\omega_S,\hat{\Omega}), 
\end{align}
and similarly for the anti-Stokes process. Here we have defined the anisotropic photonic density of states along a given direction $\hat{\Omega}$ and at a given frequency $\omega$ as
\begin{align}\label{eq:AIDoS}
D(\omega,\hat{\Omega}) \equiv \sum_\alpha \int_{\text{FBZ}} \frac{d\mb{k}}{(2\pi)^3}\delta(\omega-\omega_\alpha(\mb{k})) \delta(\hat{\Omega}-\hat{\Omega}_\alpha(\mb{k})).
\end{align}
The second delta function in the above expression selects only states around the direction of interest $\hat{\Omega}$. The total transition rate can now be written as
\begin{align}
W_S&=\int \left (\frac{dW_{S}}{d\Omega}\right ) d\Omega \notag \\
	&= n_i (n_0+1)\int |\mathcal{M}_{S}(\hat{\Omega})|^2 D(\omega_S,\hat{\Omega}) d\Omega, \notag
\end{align}
and similarly for the anti-Stokes process.

\begin{figure*}[!ht]
\begin{center}
\includegraphics[trim=0cm 0cm 0cm 0cm, clip, width=0.8\textwidth]{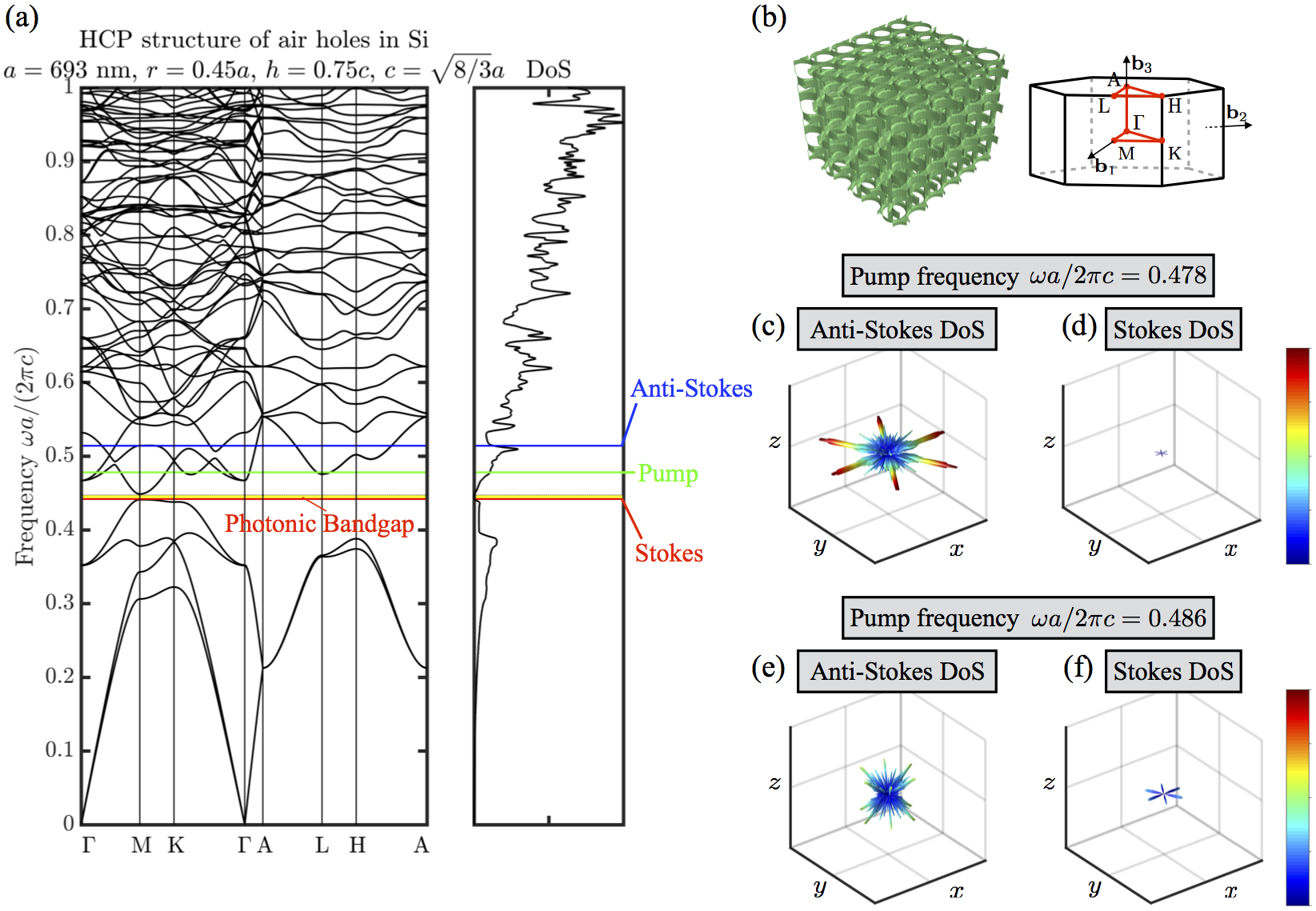}
\caption{Examples of anisotropic photonic DoS for an HCP photonic crystal of air cylinders of radius $r=0.45a$ and height $h=0.75c$ patterned in silicon, where $a$ and $c=\sqrt{8/3}a$ are the lattice parameters. \textbf{(a)} Band structure and the isotropic DoS of the system. \textbf{(b)} Structure of the photonic crystal and the $k$-point path in the Brillouin zone of the photonic crystal. We set the pump at $\omega a/2\pi c=0.478$, and use the zone-center phonon energy for silicon (519 cm$^{-1}$), corresponding to a non-dimensional frequency shift of 0.036 here ($a=693$ nm). This leads to an anti-Stokes frequency at $\omega a/2\pi c=0.514$, and a Stokes frequency at $\omega a/2\pi c=0.442$. If the pump is shifted to $\omega a/2\pi c=0.486$, the anti-Stokes frequency is shifted to $\omega a/2\pi c=0.522$, and the Stokes frequency to $\omega a/2\pi c=0.450$.  Computed anisotropic DoS \textbf{(c)} at the anti-Stokes frequency and \textbf{(d)} at the Stokes frequency when pumped at $\omega a/2\pi c=0.478$, and \textbf{(e)} at the anti-Stokes frequency and \textbf{(f)} at the Stokes frequency when pumped at $\omega a/2\pi c=0.486$. All the values of DoS are normalized by the maximum value among the four DoS plots. For anisotropic DoS at multiple frequencies along the band structure, see the supplementary video.}
\label{fig:BandHCP}
\end{center}
\end{figure*}

In carrying out the calculation of the anisotropic DoS in Eq. (\ref{eq:AIDoS}), we can rewrite $\hat{\Omega}$ in terms of polar and azimuthal angles and use the relation
\begin{align}
\delta(\hat{\Omega}-\hat{\Omega}_\alpha(\mb{k}))=\delta(\cos\theta-\cos\theta_\alpha(\mb{k}))\cdot\delta(\phi-\phi_\alpha(\mb{k})) \notag
\end{align}
to evaluate the delta function. However, writing $d\Omega=\sin\theta d\theta d\phi$ may result in uneven spacing for $d\Omega$ at different polar angles. Therefore in our calculation we rewrite Eq. (\ref{eq:AIDoS}) as
\begin{align}\label{eq:AIDoSC}
D(\omega,\hat{\Omega}) \propto \sum_\alpha \int_{\text{FBZ}} \frac{d\mb{k}}{(2\pi)^3}\delta(\omega-\omega_\alpha(\mb{k})) \delta(\hat{\Omega}\cdot\hat{\Omega}_\alpha(\mb{k})-1).
\end{align}
This change in argument in the delta function will introduce an additional constant factor to the resulting DoS. This constant factor can be eliminated by demanding that the integral of the anisotropic DoS in Eq. (\ref{eq:AIDoSC}) over entire solid angle gives the total density of states
\begin{align}
D(\omega)=\int D(\omega,\hat{\Omega}) d\Omega&=\sum_{\alpha}\int_{\text{FBZ}}\frac{d\mathbf{k}}{(2\pi)^3}\delta(\omega-\omega_\alpha(\mathbf{k})).
\end{align}

With the above expression of anisotropic DoS, we can now evaluate how the photonic DoS varies in space for our particular systems. We choose a diamond-structured photonic crystal and a hexagonal close-packed (HCP) photonic crystal for our DoS calculation. Photonic crystal structures that are similar to our HCP crystal have been discussed in \cite{joannopoulos2011photonic6} and fabricated in \cite{qi2004three} by stacking two dimensional layers. However, the structures reported in these references are the $ABCABC\cdots$ stacking, which effectively converts the photonic crystal into the face-centered cubic structure. Here, our structure is the $ABAB\cdots$ stacking and has the symmetry of the HCP structure.

For the diamond photonic crystal (Fig. \ref{fig:BandDia}), air spheres of radius $r=0.3a$ are arranged in diamond structure and patterned into crystalline silicon, where $a$ is the lattice constant of the photonic crystal lattice. For the HCP photonic crystal (Fig. \ref{fig:BandHCP}), we put air cylinders of radius $r=0.45a$ and height $h=0.75c$ at the corresponding atomic sites of an HCP crystal, where $a$ and $c=\sqrt{8/3}a$ are the lattice parameters for the HCP photonic crystal lattice.

The band structure calculation is carried out using the MIT Photonic-Bands package (MPB) \cite{Johnson2001:mpb}. We use a $24\times 24 \times 24$ grid to discretize the unit cell of both diamond and HCP photonic crystals. In the DoS calculation for the diamond photonic crystal, we use 1,110 equally-spaced $k$-points in the irreducible Brillouin zone, and a total of 42,931 $k$-points are used over the first Brillouin zone for the anisotropic DoS calculation. For the HCP photonic crystal, we use 994 equally-spaced $k$-points to discretize the irreducible Brillouin zone. The total number of $k$-points in this case is 22,439 over the first Brillouin zone. This choice of number of $k$-points allows at least 45 $k$-points on average to be around a differential solid angle in the calculation $d\Omega=4\pi/10,000$.

As examples of actual experiments, for both photonic crystals we choose two sets of pump frequencies and plot the anisotropic DoS at the anti-Stokes and Stokes frequencies corresponding to the Raman shift of the triply degenerate optical phonon modes of silicon (one LO and two TO modes, 519 cm$^{-1}$). For the diamond photonic crystal, when the system is pumped at $\omega a/2\pi c=0.594$ the Stokes frequency lies in the complete photonic band gap and thus there are no states available (Fig. \ref{fig:BandDia}d). The anti-Stokes frequency in this case is at $\omega a/2\pi c=0.639$ and the corresponding anisotropic DoS has the anisotropic shape shown in Fig. \ref{fig:BandDia}c. The peaks in the DoS completely reflect the band structure in Fig. \ref{fig:BandDia}a. This can be seen more clearly if we shift the pump to $\omega a/2\pi c=0.625$, leading to an anti-Stokes frequency at $\omega a/2\pi c=0.670$ (Fig. \ref{fig:BandDia}e) and a Stokes frequency at $\omega a/2\pi c=0.580$ (Fig. \ref{fig:BandDia}f). Now the Stokes frequency lies at the upper edge of the photonic band gap, and there are only states along the L direction in the band diagram. Thus, in the anisotropic DoS plot we see only 8 peaks corresponding to the 8 equivalent L points in the first Brillouin zone.

In comparison to the diamond photonic crystal, the DoS of the HCP photonic crystal exhibits more anisotropic behavior, particularly along the $[\textbf{001}]$ direction, as shown in Fig. \ref{fig:BandHCP}. In this structure, there is a large band gap for the out-of-plane modes but only a narrow in-plane band gap. Fig. \ref{fig:BandHCP}c and \ref{fig:BandHCP}d show the anti-Stokes DoS and the Stokes DoS when the system is pumped at $\omega a/2\pi c=0.478$. The Stokes frequency lies around the bottom edge of the band gap, and have a negligible photonic DoS. The anti-Stokes DoS, on the other hand, has large peaks along the M direction, corresponding to the low-group velocity bands around the M point.

\section{Raman selection rules and and Raman efficiency with anisotropic DoS}\label{sec:RSR}

\begin{figure*}[!ht]
\begin{center}
\includegraphics[trim=0cm 0cm 0cm 0cm, clip, width=0.65\textwidth]{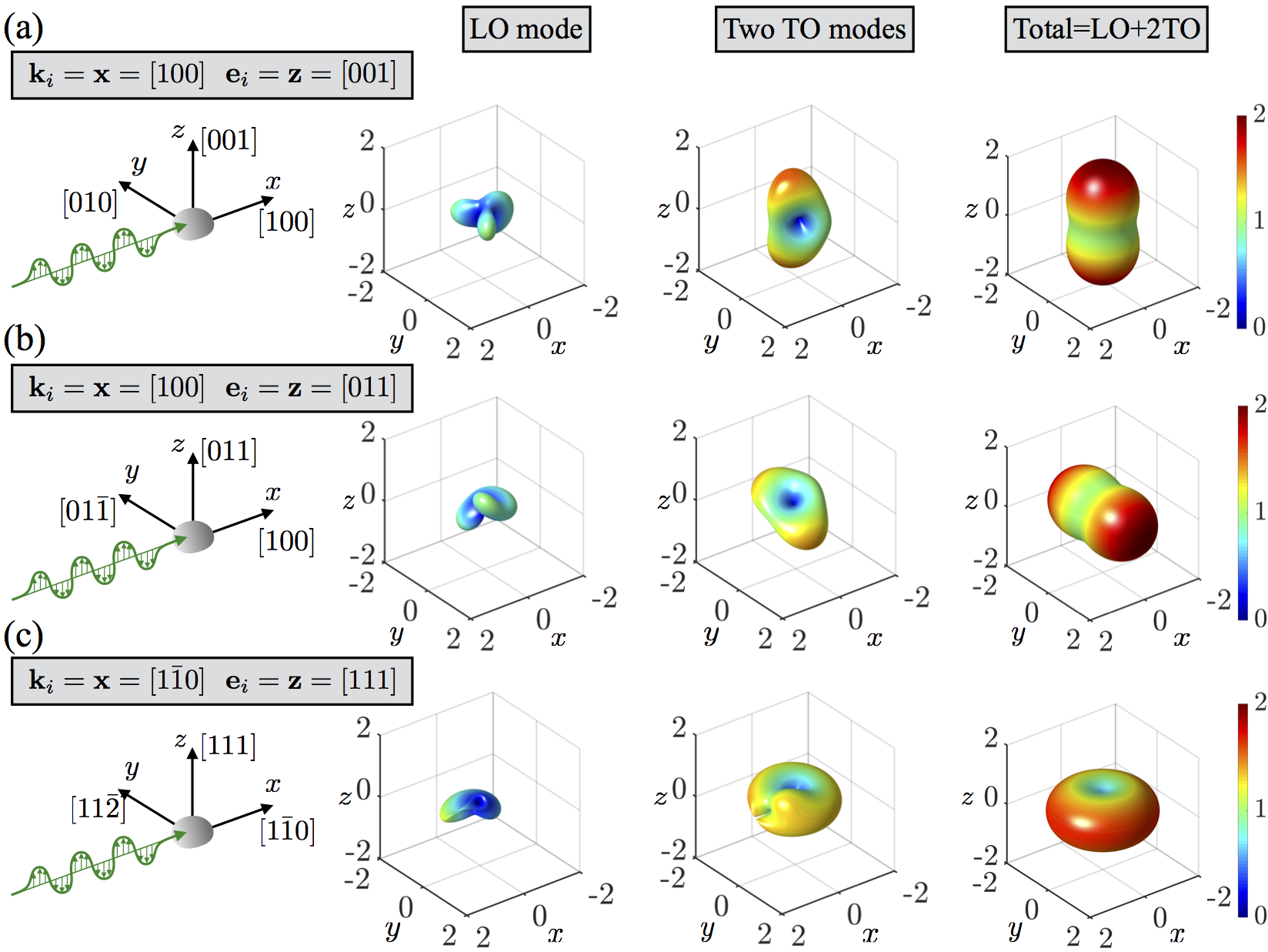}
\caption{Raman scattering pattern from Raman selection rules for different crystal orientations of silicon. In these plots we fix the incident wave vector in the $x$-direction and the polarization in the $z$-direction and rotate the material crystal. \textbf{(a)} $\mb{k}_i=\mb{x}=[100]$ and $\mb{e}_i=\mb{z}=[001]$,
\textbf{(b)} $\mb{k}_i=\mb{x}=[100]$ and $\mb{e}_i=\mb{z}=[011]$,
\textbf{(c)} $\mb{k}_i=\mb{x}=[1\bar{1}0]$ and $\mb{e}_i=\mb{z}=[111]$. The directions in Miller indices are with respect to the rotated material crystal, the actual $\mb{x}$ and $\mb{z}$ directions are fixed in all these plots. All the values are normalized to units of $|d|^2$, where $d$ is the only independent Raman tensor element for silicon.}
\label{fig:LOTO}
\end{center}
\end{figure*}

The Raman scattering process in a bulk homogeneous medium can be evaluated using Eqs. (\ref{eq:TRSI}) and (\ref{eq:TRASI}). The calculation of the matrix element $\mathcal{M}$ in Eqs. (\ref{eq:TRSI}) and (\ref{eq:TRASI}) can be carried out from first principles \cite{deinzer2002raman,lazzeri2003first}. An easier approach is to use the semi-empirical formulas for the Raman scattering efficiency per unit length per unit solid angle \cite{aggarwal2011measurement,chen2015raman,wagner1983absolute} as follows:
\begin{align}\label{eq:dSdOS}
\frac{d S_S}{d\Omega}
	=\left (\frac{\omega_S}{c} \right )^4
	\frac{N\hbar}{M\omega_0}(n_0+1)
	|R(\hat{\Omega})|^2
\end{align}
for Stokes scattering, and
\begin{align}\label{eq:dSdOAS}
\frac{d S_{AS}}{d\Omega}
	=\left (\frac{\omega_{AS}}{c} \right )^4
	\frac{N\hbar}{M\omega_0}n_0
	|R(\hat{\Omega})|^2
\end{align}
for anti-Stokes scattering. Here $N$ is the number of unit cells per unit volume, $M$ is the atomic mass, and $\hbar$ is the reduced Planck constant. The Raman tensor $R(\hat{\Omega})$ is directly proportional to the summation over the matrix elements over intermediate virtual electronic states $\mathcal{M}$ in the previous section. However, when written in this form, the independent elements in $R(\hat{\Omega})$ can be directly measured in experiments \cite{loudon1963theory,loudon1964raman,yu1996fundamentals}.
We also note that the Raman scattering efficiency is related to the Raman transition rate by the relation \cite{loudon1963theory}
\begin{align}
\frac{d S}{d\Omega}= \frac{n}{n_i c}\frac{d W}{d\Omega},\notag
\end{align}
where $n$ is the refractive index of the medium, and $n_i$ is the number of incident photons. Before we apply the anisotropic photonic DoS to the Raman scattering efficiencies, we first examine how the Raman tensor (or the material crystal) affects the Raman scattering intensity.

For incident light propagating in the direction $\mb{k}_i$ with polarization $\mb{e}_i$ scattered into the direction $\mb{k}_s$ (same as $\hat{\Omega}$ in section \ref{sec:ADoS}) with polarization $\mb{e}_s$, the total Raman intensity $|R|^2$ can be calculated through \cite{yu1996fundamentals}
\begin{align}\label{eq:RI}
|R|^2=|\mb{e}_i^T\cdot R_{\boldsymbol{\xi}} \cdot \mb{e}_s|^2,
\end{align}
where $R_{\boldsymbol{\xi}}$ is the Raman tensor for a phonon polarized in the $\boldsymbol{\xi}$ direction (the direction of atomic displacement) and the superscript ``$T$'' denotes the transpose of the vector. The Raman intensity for this scattering geometry will depend on the choice of $\mb{e}_s$. However, there is always another possible scattered photon state with polarization $\mb{e}_s'$ that is perpendicular to both $\mb{k}_s$ and $\mb{e}_s$. The total contribution to the Raman intensity from these two possible photon states is given by the sum $|\mb{e}_i^T\cdot R_{\boldsymbol{\xi}} \cdot \mb{e}_s|^2+|\mb{e}_i^T\cdot R_{\boldsymbol{\xi}} \cdot \mb{e}_s'|^2$, which is the square of the projection of the vector $(\mb{e}_i^T\cdot R_{\boldsymbol{\xi}})$ onto a plane perpendicular to $\mb{k}_s$. This is thus independent of how the two photon states are decomposed (the choice of $\mb{e}_s$ and $\mb{e}_s'$), and only depends on the direction of $\mb{k}_s$, as we mentioned in section \ref{sec:ADoS}. 

The form of the Raman tensor $R_{\boldsymbol{\xi}}$ depends on the symmetry of the corresponding phonon mode \cite{loudon1964raman}. For diamond-structured crystals ($O_h$ point group for zone center phonons) and zinc-blende-type crystals ($T_d$ point group for zone center phonons), the matrix representation of the Raman tensor for all three Raman-active phonon modes (one LO, two TO) takes the form
\begin{align}\label{eq:RT}
	R_x = \begin{pmatrix}
		0 & 0 & 0\\
		0 & 0 & d\\
		0 & d & 0\\
	\end{pmatrix},\ 
	R_y = \begin{pmatrix}
		0 & 0 & d\\
		0 & 0 & 0\\
		d & 0 & 0\\
	\end{pmatrix},\ 
	R_z = \begin{pmatrix}
		0 & d & 0\\
		d & 0 & 0\\
		0 & 0 & 0\\
	\end{pmatrix},
\end{align}
for phonons polarized along $[100]$, $[010]$, and $[001]$, respectively. Here, $d$ is the only independent Raman tensor element for these crystal structures. For a polar material (zinc-blende crystal), $d$ takes different values for LO and TO  due to the phonon-polariton effect \cite{yu1996fundamentals}. For a non-polar material such as silicon, where the LO and the two TO modes are triply degenerate, $d$ takes the same value for all these phonon modes. If the phonons are polarized in a more arbitrary direction $\boldsymbol{\xi}$, $R_{\boldsymbol{\xi}}$ will be the corresponding linear combination of the $R_x$, $R_y$, and $R_z$ matrices.

In our calculations for the selected system (crystalline silicon), we fix the incident light propagation to be along the $x$-axis $(\mb{k}_i=\mb{x}$), with a $z$-polarization of the $\mb{E}$ field ($\mb{e}_i=\mb{z}$), and allow the scattered wave vector $\mb{k}_s$ to be in any direction within the entire $4\pi$ solid angle. For each $\mb{k}_s$, we include the contributions to the Raman intensity from two possible photon polarizations that are perpendicular to the $\mb{k}_s$ direction. The wave vector $\mb{q}$ of the phonon participating in the scattering process for a selected pair of $\mb{k}_i$ and $\mb{k}_s$ is obtained from momentum conservation $\mb{q}=\mb{k}_i-\mb{k}_s$ (for Stokes scattering). The direction of phonon polarization $\boldsymbol{\xi}$ is simply $\mb{q}$ for  the LO phonon mode, or is two orthogonal unit vectors perpendicular to $\mb{q}$ for the two TO phonon modes, respectively. The corresponding Raman scattering intensity from the Raman selection rules is then evaluated through Eq. (\ref{eq:RI}).

When the material crystal is rotated with respect to a fixed incident polarization, the Raman tensor for the rotated crystal is related to the original Raman tensor by a unitary transformation
\begin{align}\label{eq:EUT}
R_{\boldsymbol{\xi}}'=UR_{\boldsymbol{\xi}}U^{T},
\end{align}
where $U$ is the Euler rotation matrix for the given rotation. Here we note that fixing the incident polarization while rotating the material is equivalent to rotating the incident polarization while fixing the material. Throughout this paper we will use the former operation.

In Fig. \ref{fig:LOTO} we show the Raman scattering intensity $|R(\Omega)|^2$ from the Raman selection rules described above for crystalline silicon. Note that we have not included the effect of anisotropic DoS and all the values are normalized by $|d|^2$. Here we choose three different orientations in which either the $[001]$, $[011]$, or $[111]$ direction coincides with the incident polarization ($\mb{e}_i=\mb{z}$) axis. The rotation matrices (corresponding to Eq. (\ref{eq:EUT})) for the $\mb{z}=[011]$ and $\mb{z}=[111]$ orientations are given by 
\begin{align}
U_{[011]}=\begin{pmatrix}
		1 & 0 & 0\\
		0 & 1/\sqrt{2} & -1/\sqrt{2}\\
		0 & 1/\sqrt{2} &1/\sqrt{2}\\
	\end{pmatrix}, \notag
\end{align}
and
\begin{align}
U_{[111]}=\begin{pmatrix}
		1/\sqrt{2} & -1/\sqrt{2} & 0\\
		1/\sqrt{6} & 1/\sqrt{6} & -2/\sqrt{6}\\
		1/\sqrt{3} & 1/\sqrt{3} &1/\sqrt{3}\\
	\end{pmatrix}, \notag
\end{align}
respectively. Silicon wafers are typically manufactured with three different choices of the out-of-plane axis, specifically the $[001]$, $[011]$, or $[111]$ axis of the crystal structure. It is thus natural to choose these out-of-plane axes as the directions of incident polarization for light, assuming an in-plane waveguiding structure. In the next section we will further allow the material crystal to rotate about this axis (the $\mb{e}_i=\mb{z}$ axis), which can be though as rotating the wafer, or chip, in a real experimental setup.

From Fig. \ref{fig:LOTO} we see that when the material crystal is rotated with respect to a fixed incident polarization, the pattern of Raman intensity is also rotated correspondingly. For the $[111]$ orientation, the shape of the total scattering pattern also changes. However, we note that the total Raman intensity integrated over all directions remains unchanged for all cases.

\vspace{10pt}

We can now include the effect of anisotropic DoS to the Raman scattering efficiencies. To do so, we first note that
Eqs. (\ref{eq:dSdOS}) and (\ref{eq:dSdOAS}) are obtained for scattering processes in free space, and the photonic DoS in vacuum $D_0(\omega)$ (Eq. (\ref{eq:IsoDoS})) is assumed. Following our treatment in the previous section to replace the vacuum DoS $D_0(\omega)$ by the actual anisotropic DoS $D(\omega,\hat{\Omega})$ and integrate over the entire solid angle, we obtain the total Raman scattering efficiency per unit length \cite{chen2015raman}:
\begin{align}\label{eq:SS}
S_{S}
	=\left (\frac{\omega_S}{c} \right )^4
	\frac{N\hbar}{M\omega_0}(n_0+1)
	\int_{\Omega}\frac{D(\omega_{S},\hat{\Omega})}{D_0(\omega_{S})}
	|R(\hat{\Omega})|^2d\Omega, 
\end{align}
\begin{align}\label{eq:SAS}
S_{AS}
	=\left (\frac{\omega_{AS}}{c} \right )^4
	\frac{N\hbar}{M\omega_0}n_0
	\int_{\Omega}\frac{D(\omega_{AS},\hat{\Omega})}{D_0(\omega_{AS})}
	|R(\hat{\Omega})|^2d\Omega.
\end{align}
From Eqs. (\ref{eq:SS}) and (\ref{eq:SAS}), we see that the term in the integral has two of degrees of freedom (in $R(\hat{\Omega})$ and $D(\omega,\hat{\Omega})$) and can be engineered to obtain the best scattering efficiency for a cooling process. In the next section we will apply this principle to optimize the Raman laser cooling process.

{In practice, one usually collects scattered light within a finite solid angle given by the experimental apparatus. The integral in } Eqs. (\ref{eq:SS}) and (\ref{eq:SAS}) {are then integrated over the solid angle in which scattered light is collected.}

\section{Optimization of figure of merit for Raman cooling}\label{sec:FoM}

Now that we know the analytical forms of both the anisotropic photonic DoS and Raman intensity, we can optimize the Raman scattering efficiencies in Eqs. \ref{eq:SS} and \ref{eq:SAS} for laser cooling. We first recall that the minimum requirement to achieve net cooling via Raman scattering from a purely spontaneous process is given by \cite{chen2015raman}
\begin{align}\label{eq:RC}
\alpha<\frac{\omega_0}{\omega_{AS}}S_{AS}-\frac{\omega_0}{\omega_{S}}S_{S},
\end{align}
where $\alpha$ is the absorption coefficient (per unit length) of the material. We note that in the above cooling condition, it is assumed that the scattered light can escape the system with 100 \% efficiency. Without further knowledge of the absorption coefficient, the most import parameter for Raman cooling is the right hand side of Eq. (\ref{eq:RC}), which we define as the figure of merit 
\begin{align}\label{eq:FoM}
\text{FoM}\equiv \frac{\omega_0}{\omega_{AS}}S_{AS}-\frac{\omega_0}{\omega_{S}}S_{S}.
\end{align}
The optimization of a Raman cooling process is therefore reduced to maximizing the FoM.

In the calculation of $S_S$ and $S_{AS}$ (Eqs. \ref{eq:SS} and \ref{eq:SAS}) for our particular systems, we use the properties of crystalline silicon, $M= 28.09$ amu, $N=2.5\times 10^{22}$ cm$^{-3}$, $|d|=1.9\times 10^{-15}$ cm$^{2}$ and the phonon energy is set to 519 cm$^{-1}$ \cite{aggarwal2011measurement} corresponding to the triply degenerate zone-center phonon modes (one LO and two TO modes).

We use the two photonic crystals discussed in section \ref{sec:ADoS} as our engineered DoS systems. Here we choose $a=861$ nm for the diamond photonic crystal such that when pumped at around 207 THz (1448 nm exhibits extremely low optical absorption for undoped silicon \cite{green2008self}) the Stokes sideband lies within the photonic band gap.

The choice of lattice constant for the HCP photonic crystal is more crucial than that of the diamond photonic crystal.
Without the knowledge of the photonic DoS, one would expect that a good choice of maximizing the FoM would have the Stokes frequency inside the band gap. However, by observing the DoS in Fig. \ref{fig:BandHCP}a, we see that if the Stokes sideband is moved into the band gap, the anti-Stokes sideband will be at around a local DoS minimum, which significantly reduces the FoM. We therefore choose $a=693$ nm, which places the Stokes frequency at the bottom edge of the photonic band gap and notably gas nonzero DoS (Fig. \ref{fig:BandHCP}d) and simultaneously place the anti-Stokes at a DoS peak when pumped at 207 THz.

\begin{figure*}[ph]
\begin{center}
\includegraphics[trim=0cm 0cm 0cm 0cm, clip, width=0.63\textwidth]{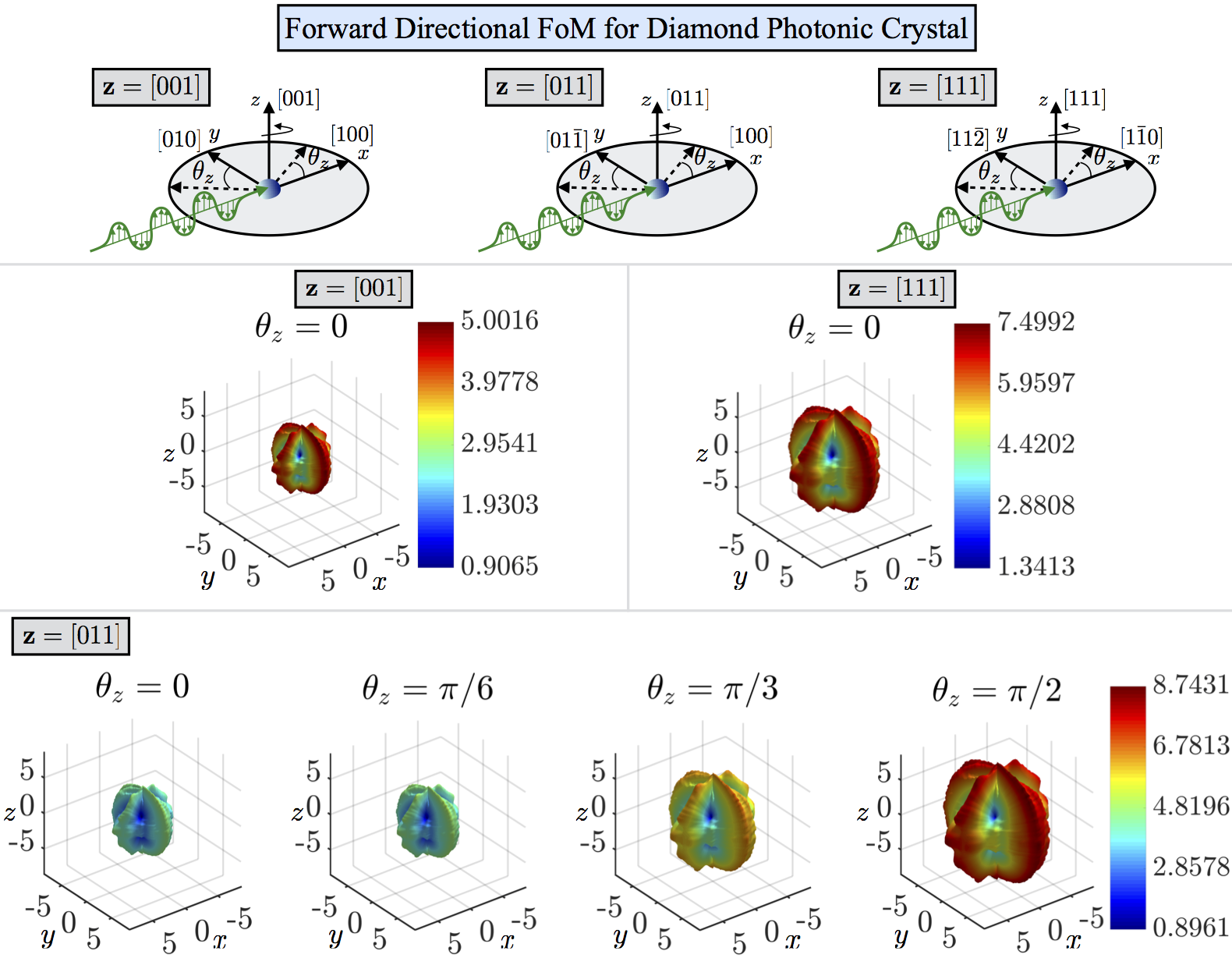}
\caption{Forward Directional Figure of Merit (FoM) over a solid angle of $\Omega=$0.77 sr around the positive $\mb{x}$ direction (see text) for total phonon scattering with the diamond photonic crystal. The scattering geometries and material crystal orientations are shown in the top-left corner. The incident light has a wave vector $\mb{k}_i=\mb{x}$ and polarization $\mb{e}_i=\mb{z}$ with respect to the fixed coordinate system ($xyz$). We allow the $[\textbf{001}]$ axis of the photonic crystal to take any direction within $4\pi$ solid angle in our fixed coordinate system, which corresponds to a direction on the 3D surface plots. The distance of a point on the surface to the origin represents the value of FoM at this particular photonic crystal orientation. All the axes are in units of $10^{-9}$ cm$^{-1}$.}
\label{fig:FDFoMDia}
\vspace{10pt}
\includegraphics[trim=0cm 0cm 0cm 0cm, clip, width=0.63\textwidth]{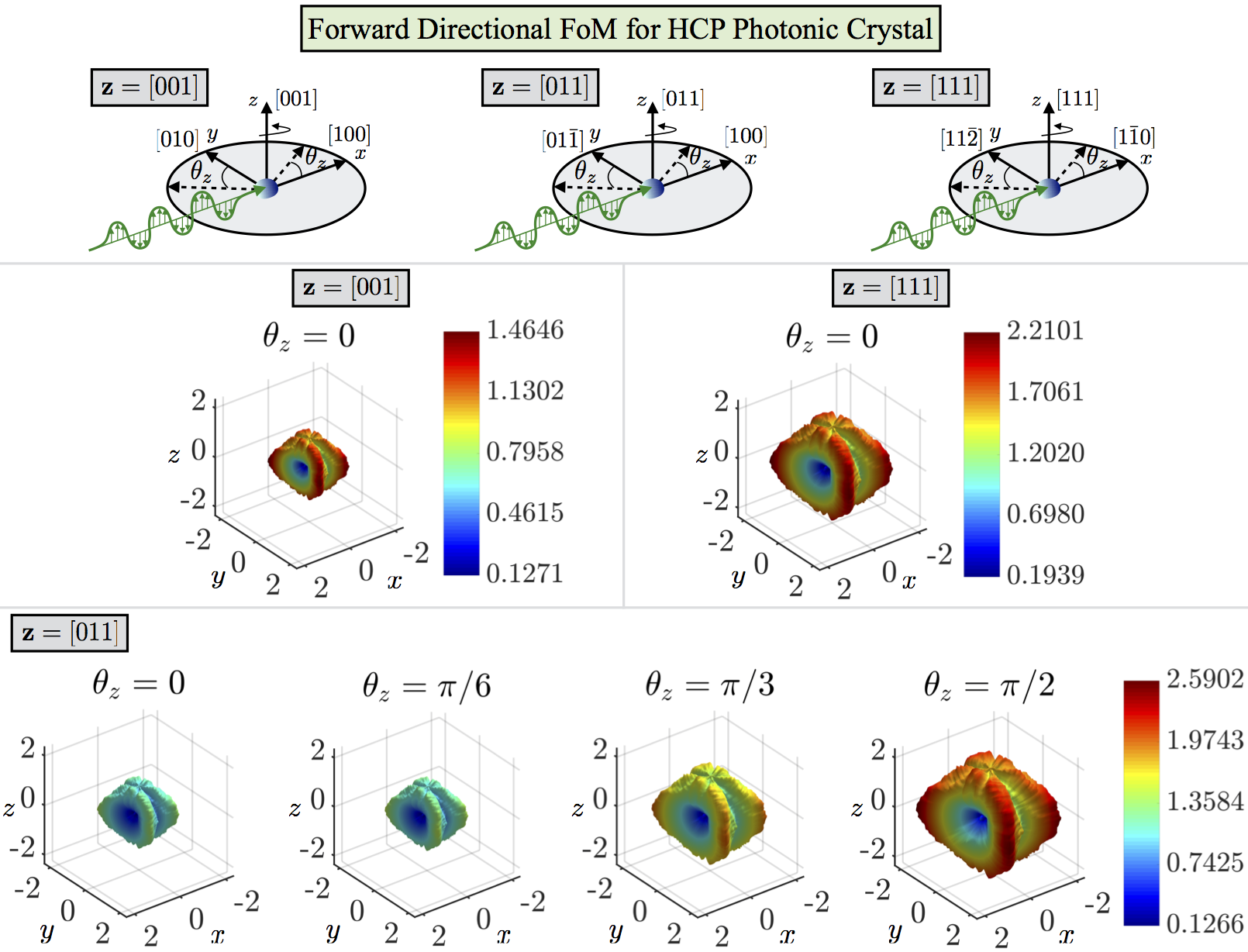}
\caption{Forward Directional Figure of Merit (FoM) over a solid angle of $\Omega=$0.77 sr around the positive $\mb{x}$ direction (see text) for total phonon scattering with the HCP photonic crystal. The scattering geometries and material crystal orientations are shown in the top-left corner. The incident light has a wave vector $\mb{k}_i=\mb{x}$ and polarization $\mb{e}_i=\mb{z}$ with respect to the fixed coordinate system ($xyz$). We allow the $[\textbf{001}]$ axis of the photonic crystal to take any direction within $4\pi$ solid angle in our fixed coordinate system, which corresponds to a direction on the 3D surface plots. The distance of a point on the surface to the origin represents the value of FoM at this particular photonic crystal orientation. All the axes are in units of $10^{-9}$ cm$^{-1}$.}
\label{fig:FDFoMHCP}
\end{center}
\end{figure*}

\begin{figure*}[ph]
\begin{center}
\includegraphics[trim=0cm 0cm 0cm 0cm, clip, width=0.63\textwidth]{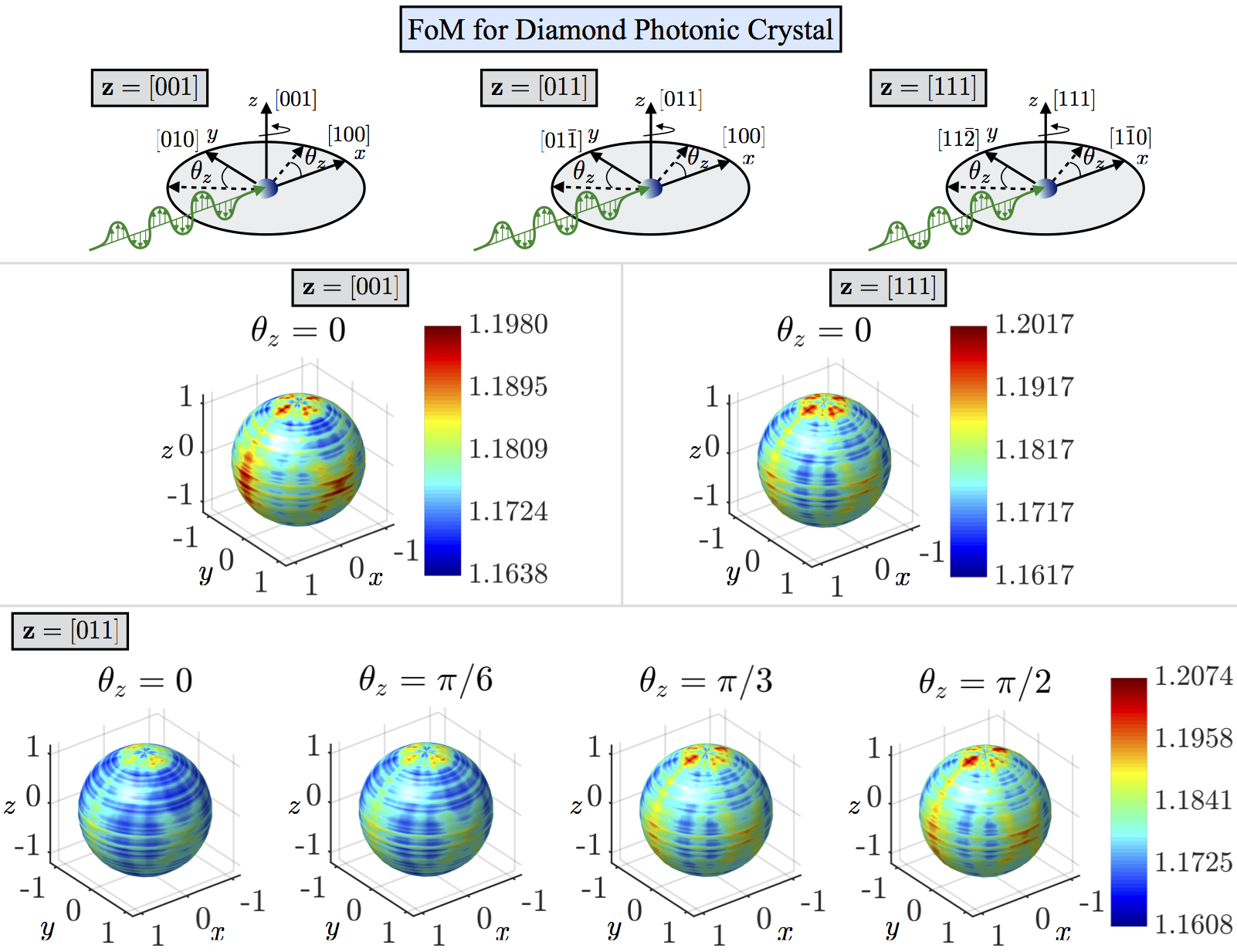}
\caption{Figure of Merit (FoM) optimization for total phonon scattering with the diamond photonic crystal. The scattering geometries and material crystal orientations are shown in the top-left corner. The incident light has a wave vector $\mb{k}_i=\mb{x}$ and polarization $\mb{e}_i=\mb{z}$ with respect to the fixed coordinate system ($xyz$). We allow the $[\textbf{001}]$ axis of the photonic crystal to take any direction within $4\pi$ solid angle with respect to our fixed coordinate system, which corresponds to a direction on the 3D surface plots. The distance of a point on the surface to the origin represents the value of FoM at this particular photonic crystal orientation. All the axes are in units of $10^{-7}$ cm$^{-1}$.}
\label{fig:FoMDia}
\vspace{10pt}
\includegraphics[trim=0cm 0cm 0cm 0cm, clip, width=0.63\textwidth]{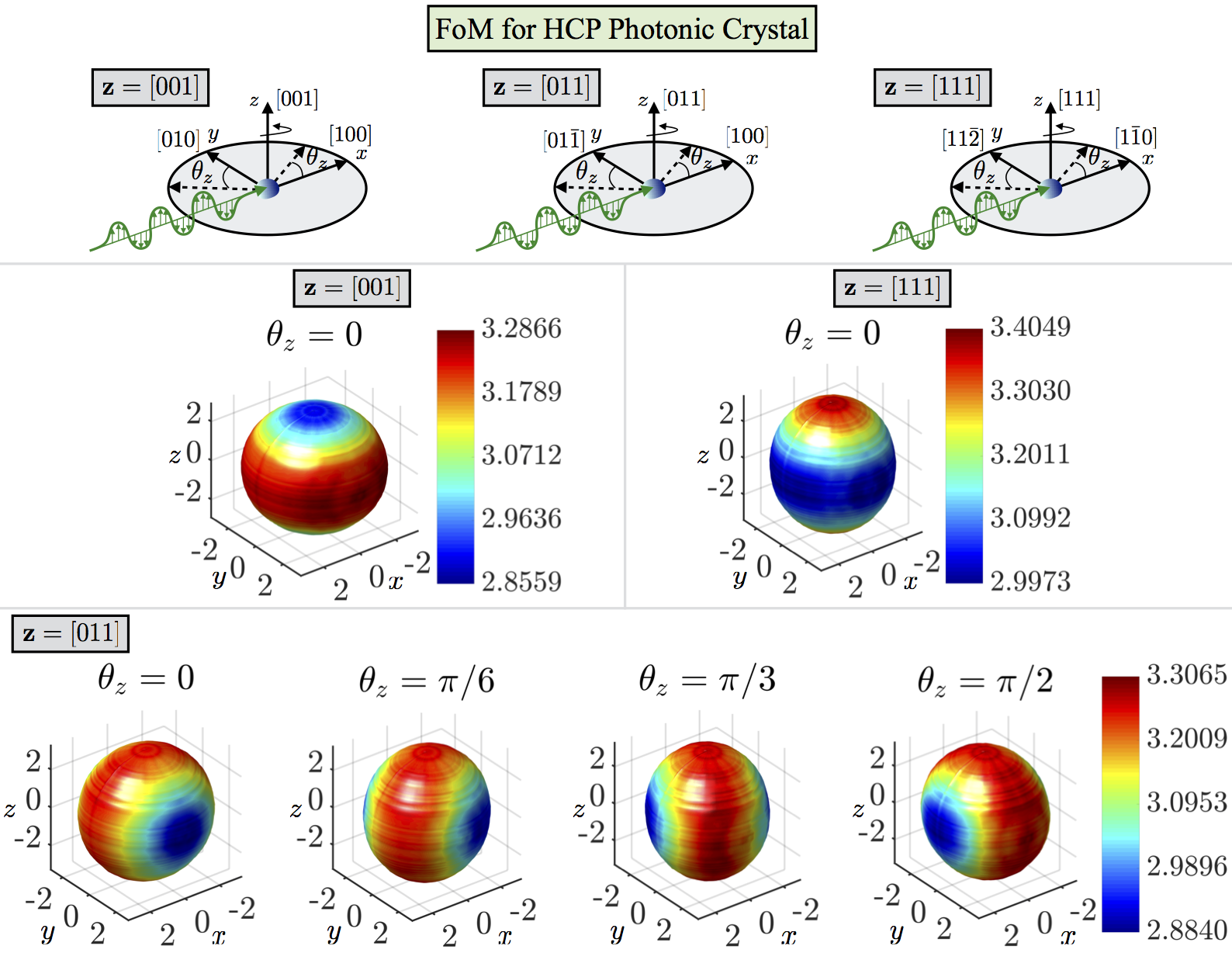}
\caption{Figure of Merit (FoM) optimization for total phonon scattering with the HCP photonic crystal. The scattering geometries and material crystal orientations are shown in the top-left corner. The incident light has a wave vector $\mb{k}_i=\mb{x}$ and polarization $\mb{e}_i=\mb{z}$ with respect to the fixed coordinate system ($xyz$). We allow the $[\textbf{001}]$ axis of the photonic crystal to take any direction within $4\pi$ solid angle with respect to our fixed coordinate system, which corresponds to a direction on the 3D surface plots. The distance of a point on the surface to the origin represents the value of FoM at this particular photonic crystal orientation. All the axes are in units of $10^{-8}$ cm$^{-1}$.}
\label{fig:FoMHCP}
\end{center}
\end{figure*}

We can now adjust the relative orientation between the photonic crystal and the material crystal to obtain the highest FoM.
To explore the optimization space, we examine cases where we set the $[001]$, $[011]$, or $[111]$ orientation of the material crystal as the $z$-axis of the fixed coordinate system, which is also the direction of incident polarization. The material crystal is now allowed to  undergo rotation on a $\mb{z}$ directed vector by an angle $\theta_z$, while keeping incident wave vector $\mb{k}_i=\mb{x}$ and polarization $\mb{e}_i=\mb{z}$ fixed (illustrated in Figs. \ref{fig:FDFoMDia}-\ref{fig:FoMHCP}). After fixing a particular material orientation, we allow the photonic crystal to take any orientation within the $4\pi$ solid angle. In our calculations, we specify the orientation of the photonic crystal using the polar and the azimuthal angles $\theta$ and $\phi$ of the $[\textbf{001}]$ axis with respect to the fixed coordinate system (not to be confused with $\theta$ and $\phi$ in section \ref{sec:ADoS}). In both diamond and HCP cases, we divide $\theta$ and $\phi$ into 100 equally-spaced segments. A total of 10,000 different orientations are used for each photonic crystal in the calculation for a chosen material orientation.

{Before discussing the effects of anisotropic DoS on Raman laser cooling, we first highlight the importance of anisotropy on the measurement of scattering. We can thus define a directional FoM, where $S_S$ and $S_{AS}$ are only integrated over the solid angle of light collection in an actual experiment. This directional FoM characterizes the fractional contribution to the total cooling by the light scattered within a given solid angle. In Figs.} \ref{fig:FDFoMDia} and \ref{fig:FDFoMHCP} {we show the forward directional FoM for the diamond and the HCP photonic crystals respectively as a function of the photonic crystal orientation for each fixed material crystal orientation (specified by a selected crystal axis as the $z$-axis, and a selected $\theta_z$). The scattering efficiencies $S_S$ and $S_{AS}$ are integrated over a cone of apex angle $\psi=1$ rad, centered around the positive $\mb{x}$ direction, which corresponds to a solid angle of $\Omega=2\pi(1-\cos\frac{\psi}{2})\approx$ 0.77 sr in the forward direction. For each 3D surface plot, the direction of a point on the surface represents the direction of the $[\textbf{001}]$ axis of the photonic crystal with respect to the fixed coordinate system, and the distance between a point on the surface to the origin indicates the value of directional FoM for the particular relative orientation. Note that for the $\mb{z}=[001]$ and the $\mb{z}=[111]$ material orientations, rotating the material crystal about the $z$-axis does not change the FoM since their total Raman scattering patterns are symmetric with respect to the $z$-axis} (Fig. \ref{fig:LOTO}a and c). Thus only one FoM surface plot is shown in Figs. \ref{fig:FDFoMDia} and \ref{fig:FDFoMHCP} for the $\mb{z}=[001]$ and the $\mb{z}=[111]$ cases.

{On the other hand, for changing photonic crystal orientation, we observe that for both the diamond and the HCP photonic crystals the forward directional FoM varies significantly. At certain photonic crystal orientations, almost no forward scattering is observed. The maximum forward directional FoM occurs for the diamond photonic crystal when the $[\textbf{001}]$ axis coincides with the $\mb{x}$ axis (and other equivalent orientations). The HCP photonic crystal, on the other hand, gives a maximum forward directional FoM when the $[\textbf{001}]$ axis lies in the $xy$ plane and makes a 69.1$^\circ$ angle from the $x$-axis.}

{
The effect of Raman selection rules can be clearly seen when we rotate the material crystal for the $\mb{z}=[011]$ case. As we increase $\theta_z$ the total Raman scattering pattern in Fig.} \ref{fig:LOTO}b also rotates about the $z$-axis, resulting in a larger forward directional FoM at $\theta_z=\pi/2$ in Figs. \ref{fig:FDFoMDia} and \ref{fig:FDFoMHCP}. {This shows that the combined knowledge of anisotropic photonic DoS and Raman selection rules is crucial here, since in an actual experiment one may not observe any scattered light if the apparatus is designed to collect light at certain direction only.}

\vspace{10pt}

We can now discuss total Raman laser cooling from scattered light in all directions. In Figs. \ref{fig:FoMDia} and \ref{fig:FoMHCP} we plot the total FoM for the diamond photonic crystal and the HCP photonic crystal respectively, where the scattering efficiencies $S_S$ and $S_{AS}$ in Eq. (\ref{eq:FoM}) are integrated over entire $4\pi$ solid angle. We see immediately that since the total FoM includes scattered light in all direction, the variation in its value is much less than the directional FoM. Overall, the FoM surfaces for the diamond photonic crystal are close to spherical, showing that the FoM for the diamond photonic crystal is relatively insensitive to changes in both material crystal orientation and photonic crystal orientation. This is due to the fact that the diamond photonic crystal belongs to the space group $O_h^7$, which has high symmetry along all directions, and the change in anisotropic DoS under rotation does not contribute too much to the integrated Raman efficiency. The largest difference in FoM among all orientations is only about 4 \%.

\begin{figure*}[!ht]
\begin{center}
\includegraphics[trim=0cm 0cm 0cm 0cm, clip, width=0.7\textwidth]{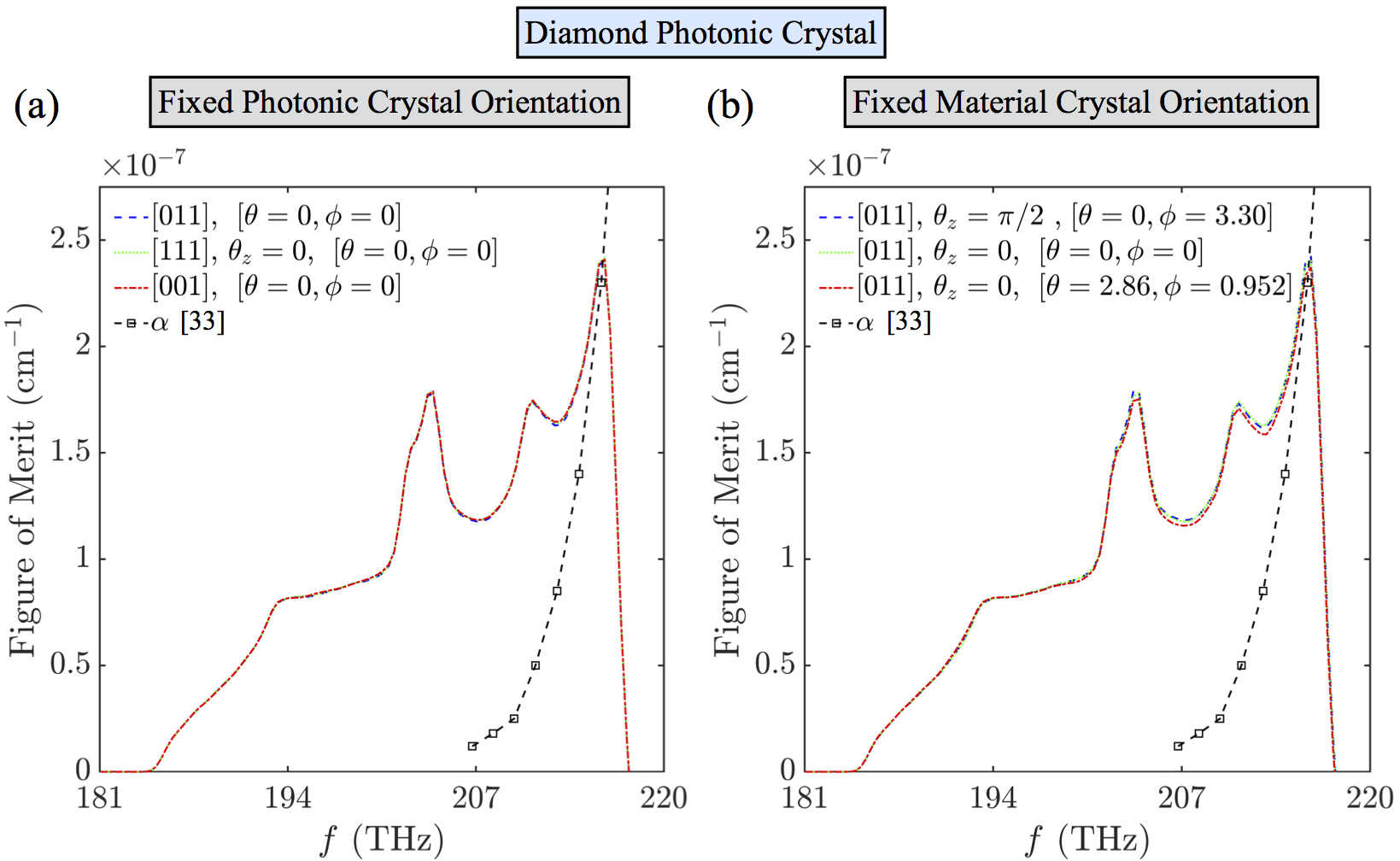}
\caption{Figure of merit of the diamond photonic crystal evaluated as a function of pump frequency for \textbf{(a)} fixed photonic crystal orientation and \textbf{(b)} fixed material crystal orientation. The experimental data for optical absorption coefficient in silicon is obtained from \cite{green2008self}.}
\label{fig:FoMvsPumpDia}
\end{center}
\end{figure*}

\begin{figure*}[!ht]
\begin{center}
\includegraphics[trim=0cm 0cm 0cm 0cm, clip, width=0.7\textwidth]{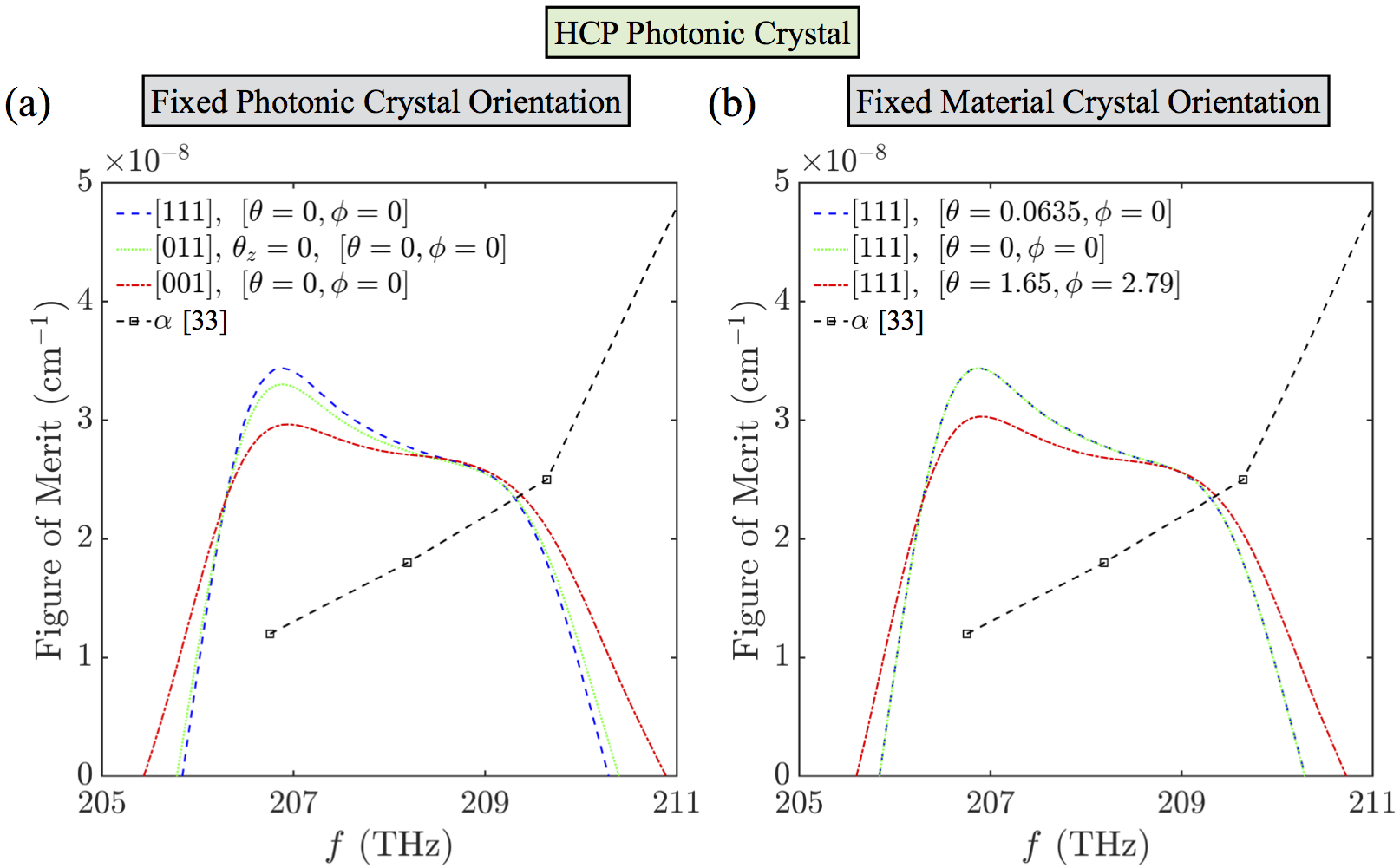}
\caption{Figure of merit of the HCP photonic crystal evaluated as a function of pump frequency for \textbf{(a)} fixed photonic crystal orientation and \textbf{(b)} fixed material crystal orientation. The experimental data for optical absorption coefficient in silicon is obtained from \cite{green2008self}.}
\label{fig:FoMvsPumpHCP}
\end{center}
\end{figure*}

The HCP photonic crystal, on the other hand, belongs to the space group $D_{6h}^4$, where properties along the out-of-plane direction behave differently from properties in the in plane directions. Thus, we see a larger difference in the FoM between cases for $[\textbf{001}]=\mb{z}$ and $[\textbf{001}]=\mb{x}$. The largest FoM for the HCP is obtained when $[111]=\mb{z}$ and $[\textbf{001}]=\mb{z}$. This can be understood from Fig. \ref{fig:BandHCP}c and Fig. \ref{fig:LOTO}c, where the large in-plane Raman scattering from the selection rules is enhanced by the anti-Stokes DoS peaks along the M directions. The largest FoM among all these orientations is up to 19 \% greater than the FoM of the worst orientation.

In these calculations for the selected examples, we use silicon as the material crystal of the system, which, as mentioned in section \ref{sec:RSR}, has the same independent Raman tensor element $d$ in Eq. (\ref{eq:RT}) for both LO and TO phonons. It is worth noting that if instead a polar material is used for Raman cooling, $d$ will be different for the LO and TO phonons. In such a case, instead of optimizing the total FoM, one can separately optimize the LO FoM or the TO FoM for the more prominent Raman mode. For example, in the excitonic resonant Raman cooling experiment reported in \cite{zhang2016resolved} (ZnTe, zinc-blende structure), the dominant phonon mode in the cooling process is the LO mode. A photonic crystal orientation that maximizes the LO FoM could be used to further enhance the cooling efficiency in the experiment.

\vspace{10pt}

In Figs. \ref{fig:FoMvsPumpDia} and \ref{fig:FoMvsPumpHCP} we plot the FoM for our particular examples as a function of pump frequency, and compare against available experimental data on optical absorption coefficients.  Here, the material crystal orientation is denoted by a choice of $z$-axis ($\mb{z}=[001]$, $\mb{z}=[011]$, or $\mb{z}=[111]$) and the angle  $\theta_z$, while the photonic crystal is denoted by $\theta$ and $\phi$, which are the polar and the azimuthal angles of the $[\textbf{001}]$ axis with respect to the fixed coordinate system. For the diamond photonic crystal, all FoM curves are almost identical and no significant change in FoM is observed. The large photonic band gap allows a wide range of frequencies (207-216 THz) for potential Raman cooling (when Eq. (\ref{eq:RC}) is satisfied). In this case, the results are similar to those calculated under the isotropic DoS assumption \cite{chen2015raman}.

We can also see that overall the FoM for the HCP photonic crystal is smaller than that of the diamond photonic crystal. This is due to the reduced dielectric material to air volume ratio in the HCP structure in order to open a band gap. However, the importance of arranging the orientation of anisotropic DoS with respect to the material crystal can be seen more clearly in Fig. \ref{fig:FoMvsPumpHCP} for the HCP case. In particular, without the knowledge of the anisotropic DoS of the system, a natural choice for fabricating such a photonic crystal would be using the most easily available $\mb{z}=[001]$ oriented wafer (red curve in Fig. \ref{fig:FoMvsPumpHCP}a). By simply using a $\mb{z}=[111]$ wafer in place of the $\mb{z}=[001]$ wafer, we can obtain an improvement of of FoM by 19 \% for a 207 THz pump.

\section{Conclusion}
We have provided the first mathematical description of how Raman scattering is influenced by anisotropic photonic DoS and Raman selection rules. Using this, we demonstrated a optimization procedure with anisotropic DoS for laser cooling by going through all possible relative crystal orientations. In the specific cases that we considered, the highly-symmetric diamond photonic crystals showed only small change in the Raman cooling FoM with respect to changes in material and photonic crystal orientations. For such photonic crystals, the isotropic DoS assumption used in \cite{chen2015raman} is sufficient to predict the range of pump frequencies for potential Raman cooling.

On the other hand, for photonic systems with lower symmetry (e.g. waveguides), the anisotropic-DoS optimization of the FoM is of greater importance. In the specific HCP photonic crystal that we considered in this paper, the difference between optimized and unoptimized FoM can be up to 19 \% and is dictated purely by a choice of crystal orientations. For such systems, isotropic DoS alone is not sufficient to predict the best available cooling efficiency of the system. The additional knowledge provided by the anisotropic DoS and the optimization procedure that examines all possible relative orientations give us the capability to identify the best orientations and frequency range for achieving Raman laser cooling.

We note that even though the 19 \% enhancement in total FoM may not be significant, the combined knowledge of both anisotropic DoS and Raman selection rules is crucial for both high- and low- symmetry photonic crystals when only directional scattering is considered. In the cases we considered, depending on the relative orientations the maximum directional FoM can be 20 times greater than the minimum (for the HCP case), which is of great importance in actual experiments.

Our formalism not only provides an approach to analyze net Raman cooling, but can also function as a method to improve the efficiency for other laser cooling methods in semiconductors \cite{zhang2013laser,zhang2016resolved} by aligning the best anisotropic DoS at the frequency of emission.

\begin{acknowledgments}
This work was funded by a US Army Research Office Grant (W911NF-15-1-0588), a US Air Force Office for Scientific Research grant (FA9550-14-1-0217), a University of Illinois Campus Research Board grant (RB15183) and a National Science Foundation grant (DMR-1555153). I.G. acknowledges support from the Illinois Scholars Undergraduate Research Program.
\end{acknowledgments}

\bibliographystyle{apsrev4-1}
\bibliography{AnisoDoSArxiv2}

\begin{thebibliography}{33}%
\makeatletter
\providecommand \@ifxundefined [1]{%
 \@ifx{#1\undefined}
}%
\providecommand \@ifnum [1]{%
 \ifnum #1\expandafter \@firstoftwo
 \else \expandafter \@secondoftwo
 \fi
}%
\providecommand \@ifx [1]{%
 \ifx #1\expandafter \@firstoftwo
 \else \expandafter \@secondoftwo
 \fi
}%
\providecommand \natexlab [1]{#1}%
\providecommand \enquote  [1]{``#1''}%
\providecommand \bibnamefont  [1]{#1}%
\providecommand \bibfnamefont [1]{#1}%
\providecommand \citenamefont [1]{#1}%
\providecommand \href@noop [0]{\@secondoftwo}%
\providecommand \href [0]{\begingroup \@sanitize@url \@href}%
\providecommand \@href[1]{\@@startlink{#1}\@@href}%
\providecommand \@@href[1]{\endgroup#1\@@endlink}%
\providecommand \@sanitize@url [0]{\catcode `\\12\catcode `\$12\catcode
  `\&12\catcode `\#12\catcode `\^12\catcode `\_12\catcode `\%12\relax}%
\providecommand \@@startlink[1]{}%
\providecommand \@@endlink[0]{}%
\providecommand \url  [0]{\begingroup\@sanitize@url \@url }%
\providecommand \@url [1]{\endgroup\@href {#1}{\urlprefix }}%
\providecommand \urlprefix  [0]{URL }%
\providecommand \Eprint [0]{\href }%
\providecommand \doibase [0]{http://dx.doi.org/}%
\providecommand \selectlanguage [0]{\@gobble}%
\providecommand \bibinfo  [0]{\@secondoftwo}%
\providecommand \bibfield  [0]{\@secondoftwo}%
\providecommand \translation [1]{[#1]}%
\providecommand \BibitemOpen [0]{}%
\providecommand \bibitemStop [0]{}%
\providecommand \bibitemNoStop [0]{.\EOS\space}%
\providecommand \EOS [0]{\spacefactor3000\relax}%
\providecommand \BibitemShut  [1]{\csname bibitem#1\endcsname}%
\let\auto@bib@innerbib\@empty
\bibitem [{\citenamefont {Pringsheim}(1929)}]{pringsheim1929zwei}%
  \BibitemOpen
  \bibfield  {author} {\bibinfo {author} {\bibfnamefont {P.}~\bibnamefont
  {Pringsheim}},\ }\href@noop {} {\bibfield  {journal} {\bibinfo  {journal}
  {Zeitschrift f{\"u}r Physik}\ }\textbf {\bibinfo {volume} {57}},\ \bibinfo
  {pages} {739} (\bibinfo {year} {1929})}\BibitemShut {NoStop}%
\bibitem [{\citenamefont {H{\"a}nsch}\ and\ \citenamefont
  {Schawlow}(1975)}]{hansch1975cooling}%
  \BibitemOpen
  \bibfield  {author} {\bibinfo {author} {\bibfnamefont {T.~W.}\ \bibnamefont
  {H{\"a}nsch}}\ and\ \bibinfo {author} {\bibfnamefont {A.~L.}\ \bibnamefont
  {Schawlow}},\ }\href@noop {} {\bibfield  {journal} {\bibinfo  {journal}
  {Optics Communications}\ }\textbf {\bibinfo {volume} {13}},\ \bibinfo {pages}
  {68} (\bibinfo {year} {1975})}\BibitemShut {NoStop}%
\bibitem [{\citenamefont {Dalibard}\ and\ \citenamefont
  {Cohen-Tannoudji}(1985)}]{dalibard1985dressed}%
  \BibitemOpen
  \bibfield  {author} {\bibinfo {author} {\bibfnamefont {J.}~\bibnamefont
  {Dalibard}}\ and\ \bibinfo {author} {\bibfnamefont {C.}~\bibnamefont
  {Cohen-Tannoudji}},\ }\href@noop {} {\bibfield  {journal} {\bibinfo
  {journal} {JOSA B}\ }\textbf {\bibinfo {volume} {2}},\ \bibinfo {pages}
  {1707} (\bibinfo {year} {1985})}\BibitemShut {NoStop}%
\bibitem [{\citenamefont {Aspect}\ \emph {et~al.}(1988)\citenamefont {Aspect},
  \citenamefont {Arimondo}, \citenamefont {Kaiser}, \citenamefont
  {Vansteenkiste},\ and\ \citenamefont {Cohen-Tannoudji}}]{aspect1988laser}%
  \BibitemOpen
  \bibfield  {author} {\bibinfo {author} {\bibfnamefont {A.}~\bibnamefont
  {Aspect}}, \bibinfo {author} {\bibfnamefont {E.}~\bibnamefont {Arimondo}},
  \bibinfo {author} {\bibfnamefont {R.~e.~a.}\ \bibnamefont {Kaiser}}, \bibinfo
  {author} {\bibfnamefont {N.}~\bibnamefont {Vansteenkiste}}, \ and\ \bibinfo
  {author} {\bibfnamefont {C.}~\bibnamefont {Cohen-Tannoudji}},\ }\href@noop {}
  {\bibfield  {journal} {\bibinfo  {journal} {Physical Review Letters}\
  }\textbf {\bibinfo {volume} {61}},\ \bibinfo {pages} {826} (\bibinfo {year}
  {1988})}\BibitemShut {NoStop}%
\bibitem [{\citenamefont {Epstein}\ \emph {et~al.}(1995)\citenamefont
  {Epstein}, \citenamefont {Buchwald}, \citenamefont {Edwards}, \citenamefont
  {Gosnell},\ and\ \citenamefont {Mungan}}]{epstein1995observation}%
  \BibitemOpen
  \bibfield  {author} {\bibinfo {author} {\bibfnamefont {R.~I.}\ \bibnamefont
  {Epstein}}, \bibinfo {author} {\bibfnamefont {M.~I.}\ \bibnamefont
  {Buchwald}}, \bibinfo {author} {\bibfnamefont {B.~C.}\ \bibnamefont
  {Edwards}}, \bibinfo {author} {\bibfnamefont {T.~R.}\ \bibnamefont
  {Gosnell}}, \ and\ \bibinfo {author} {\bibfnamefont {C.~E.}\ \bibnamefont
  {Mungan}},\ }\href@noop {} {\bibfield  {journal} {\bibinfo  {journal}
  {Nature}\ }\textbf {\bibinfo {volume} {377}},\ \bibinfo {pages} {500}
  (\bibinfo {year} {1995})}\BibitemShut {NoStop}%
\bibitem [{\citenamefont {Seletskiy}\ \emph {et~al.}(2010)\citenamefont
  {Seletskiy}, \citenamefont {Melgaard}, \citenamefont {Bigotta}, \citenamefont
  {Di~Lieto}, \citenamefont {Tonelli},\ and\ \citenamefont
  {Sheik-Bahae}}]{seletskiy2010laser}%
  \BibitemOpen
  \bibfield  {author} {\bibinfo {author} {\bibfnamefont {D.~V.}\ \bibnamefont
  {Seletskiy}}, \bibinfo {author} {\bibfnamefont {S.~D.}\ \bibnamefont
  {Melgaard}}, \bibinfo {author} {\bibfnamefont {S.}~\bibnamefont {Bigotta}},
  \bibinfo {author} {\bibfnamefont {A.}~\bibnamefont {Di~Lieto}}, \bibinfo
  {author} {\bibfnamefont {M.}~\bibnamefont {Tonelli}}, \ and\ \bibinfo
  {author} {\bibfnamefont {M.}~\bibnamefont {Sheik-Bahae}},\ }\href@noop {}
  {\bibfield  {journal} {\bibinfo  {journal} {Nature Photonics}\ }\textbf
  {\bibinfo {volume} {4}},\ \bibinfo {pages} {161} (\bibinfo {year}
  {2010})}\BibitemShut {NoStop}%
\bibitem [{\citenamefont {Seletskiy}\ \emph {et~al.}(2012)\citenamefont
  {Seletskiy}, \citenamefont {Hehlen}, \citenamefont {Epstein},\ and\
  \citenamefont {Sheik-Bahae}}]{seletskiy2012cryogenic}%
  \BibitemOpen
  \bibfield  {author} {\bibinfo {author} {\bibfnamefont {D.~V.}\ \bibnamefont
  {Seletskiy}}, \bibinfo {author} {\bibfnamefont {M.~P.}\ \bibnamefont
  {Hehlen}}, \bibinfo {author} {\bibfnamefont {R.~I.}\ \bibnamefont {Epstein}},
  \ and\ \bibinfo {author} {\bibfnamefont {M.}~\bibnamefont {Sheik-Bahae}},\
  }\href@noop {} {\bibfield  {journal} {\bibinfo  {journal} {Advances in Optics
  and Photonics}\ }\textbf {\bibinfo {volume} {4}},\ \bibinfo {pages} {78}
  (\bibinfo {year} {2012})}\BibitemShut {NoStop}%
\bibitem [{\citenamefont {Melgaard}\ \emph {et~al.}(2013)\citenamefont
  {Melgaard}, \citenamefont {Seletskiy}, \citenamefont {Di~Lieto},
  \citenamefont {Tonelli},\ and\ \citenamefont
  {Sheik-Bahae}}]{melgaard2013optical}%
  \BibitemOpen
  \bibfield  {author} {\bibinfo {author} {\bibfnamefont {S.~D.}\ \bibnamefont
  {Melgaard}}, \bibinfo {author} {\bibfnamefont {D.~V.}\ \bibnamefont
  {Seletskiy}}, \bibinfo {author} {\bibfnamefont {A.}~\bibnamefont {Di~Lieto}},
  \bibinfo {author} {\bibfnamefont {M.}~\bibnamefont {Tonelli}}, \ and\
  \bibinfo {author} {\bibfnamefont {M.}~\bibnamefont {Sheik-Bahae}},\
  }\href@noop {} {\bibfield  {journal} {\bibinfo  {journal} {Optics Letters}\
  }\textbf {\bibinfo {volume} {38}},\ \bibinfo {pages} {1588} (\bibinfo {year}
  {2013})}\BibitemShut {NoStop}%
\bibitem [{\citenamefont {Melgaard}\ \emph {et~al.}(2014)\citenamefont
  {Melgaard}, \citenamefont {Seletskiy}, \citenamefont {Polyak}, \citenamefont
  {Asmerom},\ and\ \citenamefont {Sheik-Bahae}}]{melgaard2014identification}%
  \BibitemOpen
  \bibfield  {author} {\bibinfo {author} {\bibfnamefont {S.}~\bibnamefont
  {Melgaard}}, \bibinfo {author} {\bibfnamefont {D.}~\bibnamefont {Seletskiy}},
  \bibinfo {author} {\bibfnamefont {V.}~\bibnamefont {Polyak}}, \bibinfo
  {author} {\bibfnamefont {Y.}~\bibnamefont {Asmerom}}, \ and\ \bibinfo
  {author} {\bibfnamefont {M.}~\bibnamefont {Sheik-Bahae}},\ }\href@noop {}
  {\bibfield  {journal} {\bibinfo  {journal} {Optics Express}\ }\textbf
  {\bibinfo {volume} {22}},\ \bibinfo {pages} {7756} (\bibinfo {year}
  {2014})}\BibitemShut {NoStop}%
\bibitem [{\citenamefont {Zhang}\ \emph {et~al.}(2013)\citenamefont {Zhang},
  \citenamefont {Li}, \citenamefont {Chen},\ and\ \citenamefont
  {Xiong}}]{zhang2013laser}%
  \BibitemOpen
  \bibfield  {author} {\bibinfo {author} {\bibfnamefont {J.}~\bibnamefont
  {Zhang}}, \bibinfo {author} {\bibfnamefont {D.}~\bibnamefont {Li}}, \bibinfo
  {author} {\bibfnamefont {R.}~\bibnamefont {Chen}}, \ and\ \bibinfo {author}
  {\bibfnamefont {Q.}~\bibnamefont {Xiong}},\ }\href@noop {} {\bibfield
  {journal} {\bibinfo  {journal} {Nature}\ }\textbf {\bibinfo {volume} {493}},\
  \bibinfo {pages} {504} (\bibinfo {year} {2013})}\BibitemShut {NoStop}%
\bibitem [{\citenamefont {Sheik-Bahae}\ and\ \citenamefont
  {Epstein}(2004)}]{PhysRevLett.92.247403}%
  \BibitemOpen
  \bibfield  {author} {\bibinfo {author} {\bibfnamefont {M.}~\bibnamefont
  {Sheik-Bahae}}\ and\ \bibinfo {author} {\bibfnamefont {R.~I.}\ \bibnamefont
  {Epstein}},\ }\href {\doibase 10.1103/PhysRevLett.92.247403} {\bibfield
  {journal} {\bibinfo  {journal} {Physical Review Letters}\ }\textbf {\bibinfo
  {volume} {92}},\ \bibinfo {pages} {247403} (\bibinfo {year}
  {2004})}\BibitemShut {NoStop}%
\bibitem [{\citenamefont {Sheik-Bahae}\ and\ \citenamefont
  {Epstein}(2009)}]{sheik2009laser}%
  \BibitemOpen
  \bibfield  {author} {\bibinfo {author} {\bibfnamefont {M.}~\bibnamefont
  {Sheik-Bahae}}\ and\ \bibinfo {author} {\bibfnamefont {R.~I.}\ \bibnamefont
  {Epstein}},\ }\href@noop {} {\bibfield  {journal} {\bibinfo  {journal} {Laser
  \& Photonics Reviews}\ }\textbf {\bibinfo {volume} {3}},\ \bibinfo {pages}
  {67} (\bibinfo {year} {2009})}\BibitemShut {NoStop}%
\bibitem [{\citenamefont {Ding}\ and\ \citenamefont
  {Khurgin}(2012)}]{ding2012anti}%
  \BibitemOpen
  \bibfield  {author} {\bibinfo {author} {\bibfnamefont {Y.~J.}\ \bibnamefont
  {Ding}}\ and\ \bibinfo {author} {\bibfnamefont {J.~B.}\ \bibnamefont
  {Khurgin}},\ }\href@noop {} {\bibfield  {journal} {\bibinfo  {journal} {Laser
  \& Photonics Reviews}\ }\textbf {\bibinfo {volume} {6}},\ \bibinfo {pages}
  {660} (\bibinfo {year} {2012})}\BibitemShut {NoStop}%
\bibitem [{\citenamefont {Melgaard}\ \emph {et~al.}(2016)\citenamefont
  {Melgaard}, \citenamefont {Albrecht}, \citenamefont {Hehlen},\ and\
  \citenamefont {Sheik-Bahae}}]{melgaard2016solid}%
  \BibitemOpen
  \bibfield  {author} {\bibinfo {author} {\bibfnamefont {S.~D.}\ \bibnamefont
  {Melgaard}}, \bibinfo {author} {\bibfnamefont {A.~R.}\ \bibnamefont
  {Albrecht}}, \bibinfo {author} {\bibfnamefont {M.~P.}\ \bibnamefont
  {Hehlen}}, \ and\ \bibinfo {author} {\bibfnamefont {M.}~\bibnamefont
  {Sheik-Bahae}},\ }\href@noop {} {\bibfield  {journal} {\bibinfo  {journal}
  {Scientific reports}\ }\textbf {\bibinfo {volume} {6}} (\bibinfo {year}
  {2016})}\BibitemShut {NoStop}%
\bibitem [{\citenamefont {Khurgin}(2006)}]{khurgin2006band}%
  \BibitemOpen
  \bibfield  {author} {\bibinfo {author} {\bibfnamefont {J.~B.}\ \bibnamefont
  {Khurgin}},\ }\href@noop {} {\bibfield  {journal} {\bibinfo  {journal}
  {Journal of Applied Physics}\ }\textbf {\bibinfo {volume} {100}},\ \bibinfo
  {pages} {113116} (\bibinfo {year} {2006})}\BibitemShut {NoStop}%
\bibitem [{\citenamefont {Zhang}\ \emph {et~al.}(2016)\citenamefont {Zhang},
  \citenamefont {Zhang}, \citenamefont {Wang}, \citenamefont {Kwek},\ and\
  \citenamefont {Xiong}}]{zhang2016resolved}%
  \BibitemOpen
  \bibfield  {author} {\bibinfo {author} {\bibfnamefont {J.}~\bibnamefont
  {Zhang}}, \bibinfo {author} {\bibfnamefont {Q.}~\bibnamefont {Zhang}},
  \bibinfo {author} {\bibfnamefont {X.}~\bibnamefont {Wang}}, \bibinfo {author}
  {\bibfnamefont {L.~C.}\ \bibnamefont {Kwek}}, \ and\ \bibinfo {author}
  {\bibfnamefont {Q.}~\bibnamefont {Xiong}},\ }\href@noop {} {\bibfield
  {journal} {\bibinfo  {journal} {Nature Photonics}\ }\textbf {\bibinfo
  {volume} {10}},\ \bibinfo {pages} {600} (\bibinfo {year} {2016})}\BibitemShut
  {NoStop}%
\bibitem [{\citenamefont {Bahl}\ \emph {et~al.}(2012)\citenamefont {Bahl},
  \citenamefont {Tomes}, \citenamefont {Marquardt},\ and\ \citenamefont
  {Carmon}}]{bahl2012observation}%
  \BibitemOpen
  \bibfield  {author} {\bibinfo {author} {\bibfnamefont {G.}~\bibnamefont
  {Bahl}}, \bibinfo {author} {\bibfnamefont {M.}~\bibnamefont {Tomes}},
  \bibinfo {author} {\bibfnamefont {F.}~\bibnamefont {Marquardt}}, \ and\
  \bibinfo {author} {\bibfnamefont {T.}~\bibnamefont {Carmon}},\ }\href@noop {}
  {\bibfield  {journal} {\bibinfo  {journal} {Nature Physics}\ }\textbf
  {\bibinfo {volume} {8}},\ \bibinfo {pages} {203} (\bibinfo {year}
  {2012})}\BibitemShut {NoStop}%
\bibitem [{\citenamefont {Chen}\ and\ \citenamefont
  {Bahl}(2015)}]{chen2015raman}%
  \BibitemOpen
  \bibfield  {author} {\bibinfo {author} {\bibfnamefont {Y.-C.}\ \bibnamefont
  {Chen}}\ and\ \bibinfo {author} {\bibfnamefont {G.}~\bibnamefont {Bahl}},\
  }\href@noop {} {\bibfield  {journal} {\bibinfo  {journal} {Optica}\ }\textbf
  {\bibinfo {volume} {2}},\ \bibinfo {pages} {893} (\bibinfo {year}
  {2015})}\BibitemShut {NoStop}%
\bibitem [{\citenamefont {Chen}\ \emph {et~al.}(2016)\citenamefont {Chen},
  \citenamefont {Kim},\ and\ \citenamefont {Bahl}}]{chen2016brillouin}%
  \BibitemOpen
  \bibfield  {author} {\bibinfo {author} {\bibfnamefont {Y.-C.}\ \bibnamefont
  {Chen}}, \bibinfo {author} {\bibfnamefont {S.}~\bibnamefont {Kim}}, \ and\
  \bibinfo {author} {\bibfnamefont {G.}~\bibnamefont {Bahl}},\ }\href@noop {}
  {\bibfield  {journal} {\bibinfo  {journal} {New Journal of Physics}\ }\textbf
  {\bibinfo {volume} {18}},\ \bibinfo {pages} {115004} (\bibinfo {year}
  {2016})}\BibitemShut {NoStop}%
\bibitem [{\citenamefont {Kim}\ and\ \citenamefont {Bahl}(2017)}]{kim2017role}%
  \BibitemOpen
  \bibfield  {author} {\bibinfo {author} {\bibfnamefont {S.}~\bibnamefont
  {Kim}}\ and\ \bibinfo {author} {\bibfnamefont {G.}~\bibnamefont {Bahl}},\
  }\href {\doibase 10.1364/OE.25.000776} {\bibfield  {journal} {\bibinfo
  {journal} {Opt. Express}\ }\textbf {\bibinfo {volume} {25}},\ \bibinfo
  {pages} {776} (\bibinfo {year} {2017})}\BibitemShut {NoStop}%
\bibitem [{\citenamefont {Purcell}(1946)}]{purcell1946spontaneous}%
  \BibitemOpen
  \bibfield  {author} {\bibinfo {author} {\bibfnamefont {E.~M.}\ \bibnamefont
  {Purcell}},\ }\href@noop {} {\bibfield  {journal} {\bibinfo  {journal}
  {Physical Review}\ }\textbf {\bibinfo {volume} {69}},\ \bibinfo {pages} {681}
  (\bibinfo {year} {1946})}\BibitemShut {NoStop}%
\bibitem [{\citenamefont {Gaponenko}(2002)}]{gaponenko2002effects}%
  \BibitemOpen
  \bibfield  {author} {\bibinfo {author} {\bibfnamefont {S.}~\bibnamefont
  {Gaponenko}},\ }\href@noop {} {\bibfield  {journal} {\bibinfo  {journal}
  {Physical Review B}\ }\textbf {\bibinfo {volume} {65}},\ \bibinfo {pages}
  {140303} (\bibinfo {year} {2002})}\BibitemShut {NoStop}%
\bibitem [{\citenamefont {Loudon}(1963)}]{loudon1963theory}%
  \BibitemOpen
  \bibfield  {author} {\bibinfo {author} {\bibfnamefont {R.}~\bibnamefont
  {Loudon}},\ }\href@noop {} {\bibfield  {journal} {\bibinfo  {journal}
  {Proceedings of the Royal Society of London. Series A. Mathematical and
  Physical Sciences}\ }\textbf {\bibinfo {volume} {275}},\ \bibinfo {pages}
  {218} (\bibinfo {year} {1963})}\BibitemShut {NoStop}%
\bibitem [{\citenamefont {Loudon}(1964)}]{loudon1964raman}%
  \BibitemOpen
  \bibfield  {author} {\bibinfo {author} {\bibfnamefont {R.}~\bibnamefont
  {Loudon}},\ }\href@noop {} {\bibfield  {journal} {\bibinfo  {journal}
  {Advances in Physics}\ }\textbf {\bibinfo {volume} {13}},\ \bibinfo {pages}
  {423} (\bibinfo {year} {1964})}\BibitemShut {NoStop}%
\bibitem [{\citenamefont {Yu}\ and\ \citenamefont
  {Cardona}(1996)}]{yu1996fundamentals}%
  \BibitemOpen
  \bibfield  {author} {\bibinfo {author} {\bibfnamefont {P.~Y.}\ \bibnamefont
  {Yu}}\ and\ \bibinfo {author} {\bibfnamefont {M.}~\bibnamefont {Cardona}},\
  }\enquote {\bibinfo {title} {Fundamentals of semiconductors},}\ \ (\bibinfo
  {publisher} {Springer},\ \bibinfo {year} {1996})\ Chap.~\bibinfo {chapter}
  {7}\BibitemShut {NoStop}%
\bibitem [{\citenamefont {Joannopoulos}\ \emph {et~al.}(2011)\citenamefont
  {Joannopoulos}, \citenamefont {Johnson}, \citenamefont {Winn},\ and\
  \citenamefont {Meade}}]{joannopoulos2011photonic6}%
  \BibitemOpen
  \bibfield  {author} {\bibinfo {author} {\bibfnamefont {J.~D.}\ \bibnamefont
  {Joannopoulos}}, \bibinfo {author} {\bibfnamefont {S.~G.}\ \bibnamefont
  {Johnson}}, \bibinfo {author} {\bibfnamefont {J.~N.}\ \bibnamefont {Winn}}, \
  and\ \bibinfo {author} {\bibfnamefont {R.~D.}\ \bibnamefont {Meade}},\
  }\enquote {\bibinfo {title} {Photonic crystals: molding the flow of light},}\
  \ (\bibinfo  {publisher} {Princeton university press},\ \bibinfo {year}
  {2011})\ Chap.~\bibinfo {chapter} {6}\BibitemShut {NoStop}%
\bibitem [{\citenamefont {Qi}\ \emph {et~al.}(2004)\citenamefont {Qi},
  \citenamefont {Lidorikis}, \citenamefont {Rakich}, \citenamefont {Johnson}
  \emph {et~al.}}]{qi2004three}%
  \BibitemOpen
  \bibfield  {author} {\bibinfo {author} {\bibfnamefont {M.}~\bibnamefont
  {Qi}}, \bibinfo {author} {\bibfnamefont {E.}~\bibnamefont {Lidorikis}},
  \bibinfo {author} {\bibfnamefont {P.~T.}\ \bibnamefont {Rakich}}, \bibinfo
  {author} {\bibfnamefont {S.~G.}\ \bibnamefont {Johnson}},  \emph {et~al.},\
  }\href@noop {} {\bibfield  {journal} {\bibinfo  {journal} {Nature}\ }\textbf
  {\bibinfo {volume} {429}},\ \bibinfo {pages} {538} (\bibinfo {year}
  {2004})}\BibitemShut {NoStop}%
\bibitem [{\citenamefont {Johnson}\ and\ \citenamefont
  {Joannopoulos}(2001)}]{Johnson2001:mpb}%
  \BibitemOpen
  \bibfield  {author} {\bibinfo {author} {\bibfnamefont {S.~G.}\ \bibnamefont
  {Johnson}}\ and\ \bibinfo {author} {\bibfnamefont {J.~D.}\ \bibnamefont
  {Joannopoulos}},\ }\href
  {http://www.opticsexpress.org/abstract.cfm?URI=OPEX-8-3-173} {\bibfield
  {journal} {\bibinfo  {journal} {Opt. Express}\ }\textbf {\bibinfo {volume}
  {8}},\ \bibinfo {pages} {173} (\bibinfo {year} {2001})}\BibitemShut {NoStop}%
\bibitem [{\citenamefont {Deinzer}\ and\ \citenamefont
  {Strauch}(2002)}]{deinzer2002raman}%
  \BibitemOpen
  \bibfield  {author} {\bibinfo {author} {\bibfnamefont {G.}~\bibnamefont
  {Deinzer}}\ and\ \bibinfo {author} {\bibfnamefont {D.}~\bibnamefont
  {Strauch}},\ }\href@noop {} {\bibfield  {journal} {\bibinfo  {journal}
  {Physical Review B}\ }\textbf {\bibinfo {volume} {66}},\ \bibinfo {pages}
  {100301} (\bibinfo {year} {2002})}\BibitemShut {NoStop}%
\bibitem [{\citenamefont {Lazzeri}\ and\ \citenamefont
  {Mauri}(2003)}]{lazzeri2003first}%
  \BibitemOpen
  \bibfield  {author} {\bibinfo {author} {\bibfnamefont {M.}~\bibnamefont
  {Lazzeri}}\ and\ \bibinfo {author} {\bibfnamefont {F.}~\bibnamefont
  {Mauri}},\ }\href@noop {} {\bibfield  {journal} {\bibinfo  {journal}
  {Physical Review Letters}\ }\textbf {\bibinfo {volume} {90}},\ \bibinfo
  {pages} {036401} (\bibinfo {year} {2003})}\BibitemShut {NoStop}%
\bibitem [{\citenamefont {Aggarwal}\ \emph {et~al.}(2011)\citenamefont
  {Aggarwal}, \citenamefont {Farrar}, \citenamefont {Saikin}, \citenamefont
  {Aspuru-Guzik}, \citenamefont {Stopa},\ and\ \citenamefont
  {Polla}}]{aggarwal2011measurement}%
  \BibitemOpen
  \bibfield  {author} {\bibinfo {author} {\bibfnamefont {R.}~\bibnamefont
  {Aggarwal}}, \bibinfo {author} {\bibfnamefont {L.}~\bibnamefont {Farrar}},
  \bibinfo {author} {\bibfnamefont {S.}~\bibnamefont {Saikin}}, \bibinfo
  {author} {\bibfnamefont {A.}~\bibnamefont {Aspuru-Guzik}}, \bibinfo {author}
  {\bibfnamefont {M.}~\bibnamefont {Stopa}}, \ and\ \bibinfo {author}
  {\bibfnamefont {D.}~\bibnamefont {Polla}},\ }\href@noop {} {\bibfield
  {journal} {\bibinfo  {journal} {Solid State Communications}\ }\textbf
  {\bibinfo {volume} {151}},\ \bibinfo {pages} {553} (\bibinfo {year}
  {2011})}\BibitemShut {NoStop}%
\bibitem [{\citenamefont {Wagner}\ and\ \citenamefont
  {Cardona}(1983)}]{wagner1983absolute}%
  \BibitemOpen
  \bibfield  {author} {\bibinfo {author} {\bibfnamefont {J.}~\bibnamefont
  {Wagner}}\ and\ \bibinfo {author} {\bibfnamefont {M.}~\bibnamefont
  {Cardona}},\ }\href@noop {} {\bibfield  {journal} {\bibinfo  {journal} {Solid
  State Communications}\ }\textbf {\bibinfo {volume} {48}},\ \bibinfo {pages}
  {301} (\bibinfo {year} {1983})}\BibitemShut {NoStop}%
\bibitem [{\citenamefont {Green}(2008)}]{green2008self}%
  \BibitemOpen
  \bibfield  {author} {\bibinfo {author} {\bibfnamefont {M.~A.}\ \bibnamefont
  {Green}},\ }\href@noop {} {\bibfield  {journal} {\bibinfo  {journal} {Solar
  Energy Materials and Solar Cells}\ }\textbf {\bibinfo {volume} {92}},\
  \bibinfo {pages} {1305} (\bibinfo {year} {2008})}\BibitemShut {NoStop}%
\end{thebibliography}%
\end{document}